\documentstyle[preprint,psfig]{aastex}

\def\lesssim{\mathrel{\hbox{\rlap{\hbox{\lower4pt\hbox{$\sim$}}}\hbox{$<$}}}}
\def\gtrsim{\mathrel{\hbox{\rlap{\hbox{\lower4pt\hbox{$\sim$}}}\hbox{$>$}}}}

\begin{document}
\title{Global MHD Simulations of Cylindrical Keplerian Disks}

\author{John F. Hawley}
\affil{Virginia Institute for Theoretical Astronomy,
Department of Astronomy, University of Virginia,
Charlottesville VA 22903}

\shorttitle{Cylindrical Keplerian Disks}

\begin{abstract}

This paper presents a series of global three dimensional accretion disk
simulations carried out in the cylindrical limit in which the vertical
component of the gravitational field is neglected.  The simulations use
a cylindrical pseudo-Newtonian potential, $\propto 1/(R-R_g)$, to model
the main dynamical properties of the Schwarzschild metric.  The radial
grid domain runs out to 60 $R_g$ to minimize the influence of the outer
boundary on the inner disk evolution.  The disks are initially constant
density with a Keplerian angular momentum distribution and contain a
weak toroidal or vertical field which serves as the seed for the
magnetorotational instability.  These simulations reaffirm many of the
conclusions of previous local simulations.  The MRI grows rapidly and
produces MHD turbulence with a significant Maxwell stress which drives
accretion.  Tightly-wrapped low-$m$ spiral waves are prominent.  In
some simulations radial variations in Maxwell stress concentrate gas
into rings, creating substantial spatial inhomogeneities.  As in
previous global simulations, there is a nonzero stress at the
marginally stable orbit.  The stress is smaller than seen in stratified
torus simulations, but nevertheless produces a small decline in
specific angular momentum inside the last stable orbit.  Detailed
comparisons between simulations are used to examine the effects of
various choices in computational setup.  Because the driving
instability is local, a reduction in the azimuthal computational domain
to some fraction of $2\pi$ does not create large qualitative
differences.  Similarly, the choice of either an isothermal or
adiabatic equation of state has little impact on the initial
evolution.  Simulations that begin with vertical fields have greater
field amplification and higher ratios of stress to magnetic pressure
compared with those beginning with toroidal fields.  In contrast to
MHD, hydrodynamics alone neither creates nor sustains turbulence.

\end{abstract}


\section{Introduction}

Recent increases in supercomputer performance have significantly
improved the ability to evolve the basic equations of accretion disk
structure and evolution.  These developments, along with continuing
progress in understanding the most important physical processes that
occur within accretion disks, suggest that predictive disk simulations
are a realistic goal.  Such disk simulations will be global, fully
three dimensional, and incorporate physical processes such as
magnetohydrodynamics (MHD) and radiation transport.  At present, we are
some ways from this goal; global simulations are still rather idealized
in terms of disk structure, energetics, and dynamical range.  However,
because almost any three dimensional disk simulation is relatively
novel, there remain many significant questions to be investigated even
with such simplified models.

To date there have been several global simulations of three dimensional
disks, including Armitage (1998), Matsumoto (1999), Hawley (2000),
Machida, Hayashi, \& Matsumoto (2000), Hawley \& Krolik (2001;
hereafter HK), and Armitage, Reynolds \& Chiang (2001; hereafter ARC).
Much of this work has focused on thick accretion disks.  With a
pressure scale height $H$ comparable to the disk radius $R$, the thick
disk, or accretion torus, is more easily resolved in a numerical
simulation than disks for which $H/R \ll 1$.  Matsumoto (1999) followed
the evolution of a thick torus embedded in an external vertical field,
and found significant outflow collimated along the global vertical
field lines.  Hawley (2000) considered tori containing toroidal fields
and poloidal field loops, and Machida et al.~(2000) modeled a thick
disk containing a toroidal field.  In these studies the initial field
was entirely contained within the disk and the resulting outflows were
confined to the creation of a magnetized corona.  A generic feature of
all these thick disk simulations is the presence of large amplitude
fluctuations in accretion rate, density, and other variables, in both
space and time.

At a minimum these efforts have established that the magnetorotational
instability, or MRI, (Balbus \& Hawley 1991) is just as efficacious in
thick disks as in local simulations to produce MHD turbulence and
angular momentum transport.  Thick accretion tori with initially
non-Keplerian angular momentum distributions are highly unstable.  MHD
turbulence develops rapidly and is sustained by a self-consistent
dynamo process within the disk.  The constant or near-constant specific
angular momentum distribution of the initial torus rapidly evolves to
one that is near Keplerian.  The main focus for dynamical studies
would therefore seem to be Keplerian disks, both hot (high internal
sound speed), and thin and cold (low internal sound speed).  

Global simulations have been used to investigate specific physical
issues in Keplerian accretion disk models.  HK examined the behavior of
a thick, nearly Keplerian disk model to study the accretion flow
through the radius of the marginally stable orbit ($r_{ms}$) in a
pseudo-Newtonian potential.  They found significant stress at the
location of the marginally stable orbit which creates a continuing
decline in the value of the specific angular momentum $\ell$ inside of
this point.  The equation of state was adiabatic and there was no
cooling, and the thickness of the disk remained roughly $H/R \sim 0.15$
throughout.

So far there have been fewer simulations of Keplerian disks with low
internal sound speeds corresponding to small $H/R$.  To sidestep the
difficulty of resolving both $H$ and $R$, the cylindrical disk limit
has been employed which omits vertical gravity and stratification.
Armitage (1998) and Hawley (2000) simulated a few examples of these
cylindrically-symmetric Keplerian disks.  More recently, ARC modeled
several Keplerian cylindrical disks with sound speeds corresponding to
$H/R$ of 0.08 and 0.04.  They also computed one stratified disk that
covered a limited vertical extent.  As in the work of HK, they examined
the inflow through $r_{ms}$.  Although they found a nonzero Maxwell
stress inside of $r_{ms}$, it was not large enough to alter $\ell$ to
the same extent as seen in the simulation of HK.  They emphasized
that the differences between their simulations and those of HK were
quantitative; the same dynamical effects were present, only at reduced
amplitude.  In addition to the lower initial sound speed and cylindrical
symmetry, the ARC simulations used an angular domain of $\pi/6$
compared with the full $2\pi$ simulation of HK.  It is unclear which,
if any, of these differences is the most significant in comparing the
results of HK and ARC.

Clearly, we still have only a preliminary understanding of the
specific processes that establish the precise turbulent stress levels in
disks.  An attempt to investigate this question through simulation is
further complicated by the need to distinguish between influences
created purely by the numerical details (e.g.  computational domain,
numerical resolution, initial conditions), and those determined by
physics.  In an effort to characterize the array of technical choices
in three dimensional simulations, this paper presents a series of disk
simulations that begin with a Keplerian angular momentum distribution.
These simulations are restricted to the cylindrical limit which omits
the gravitational acceleration in the $z$ direction.  The cylindrical
approximation has the distinct advantage of reducing the number of grid
zones required in $z$:  only the wavelengths of the weak-field MRI need
be resolved, not several pressure scale heights.  Here this advantage
is exploited to increase the number of grid zones devoted to the radial
extent of the disk, and its angular resolution.

Comparisons of cylindrical disks and vertically stratified disks in
Hawley (2000) provided some evidence that the cylindrical disk is a good
approximation for studying the evolution of density-averaged properties
in a disk.  Here we will be exploring the detailed properties and
limitations of these cylindrical disks in greater detail.  These
cylindrical disks will be directly comparable to the simulations of
ARC, and are the natural sequel to the local nonstratified shearing box
simulations done previously (Hawley, Gammie, \& Balbus 1995, hereafter
HGB95; 1996; Matsumoto \& Tajima 1995).  Comparison with shearing boxes
provide insight into those phenomena that are truly global versus those
that are well-described locally.  Cylindrical disk models will also
provide a baseline of results that will be compared with future
stratified Keplerian disk models.

Some of the questions that this investigation will address include:
How do these global results compare with local shearing box
simulations?  Are there effects that can be attributed to influences
from the equation of state, or which arise when compared to simulations
of tori with higher internal temperatures?  What differences are there
in disks computed on grids with angular extent less that $2\pi$?  Can a
steady state be established in at least part of the disk?  What are the
amplitudes, and the time- and length-scales for the fluctuations that
result from the MRI-induced turbulence?

The plan of this paper is as follows.  In \S 2 the computational setup
is described.  Section 3 presents the results from a series of MHD
cylindrical disk simulations.  In \S 4 a purely hydrodynamic disk is
considered for contrast.  The results and their consequences are
discussed in \S 5, and the conclusions are summarized in \S 6.  

\section{Problem Setup}

The simulations evolve the equations of ideal MHD,
\begin{equation}\label{mass}
{\partial\rho\over \partial t} + \nabla\cdot (\rho {\bf v}) =  0,
\end{equation}
\begin{equation}\label{mom}
\rho {\partial{\bf v} \over \partial t}
+ (\rho {\bf v}\cdot\nabla){\bf v} = -\nabla\left(
P + {\mathcal Q} +{B^2\over 8 \pi} \right)-\rho \nabla \Phi +
\left( {{\bf B}\over 4\pi}\cdot \nabla\right){\bf B},
\end{equation}
\begin{equation}\label{ene}
{\partial\rho\epsilon\over \partial t} + \nabla\cdot (\rho\epsilon
{\bf v}) = -(P+{\mathcal Q}) \nabla \cdot {\bf v},
\end{equation}
\begin{equation} \label{ind}
{\partial{\bf B}\over \partial t} =
\nabla\times\left( {\bf v} \times {\bf B} \right),
\end{equation}
where $\rho$ is the mass density, $\epsilon$ is the specific internal
energy, ${\bf v}$ is the fluid velocity, $P$ is the pressure,
${\mathcal Q}$ is an explicit artificial viscosity (Stone \& Norman
1992a), ${\bf B}$ is the magnetic field vector, and $\Phi$ is the
gravitational potential.  We employ cylindrical coordinates,
$(R,\phi,z)$, and work in the ``cylindrical disk'' limit which assumes
a cylindrical gravitational potential and ignores vertical
stratification.  The gravitational potential $\Phi$ is a
cylindrically-symmetric form of the Paczy\'nski \& Wiita (1980)
pseudo-Newtonian potential $\Phi = - GM/(R-R_g)$.  The equation of
state is either isothermal, $P = c_s^2 \rho$, or adiabatic,
$P=\rho\epsilon(\Gamma -1)$.  Radiation transport and losses are
omitted.  Since there is no explicit resistivity or physical viscosity,
the adiabatic gas can heat only by compression, or in nonadiabatic
heating through the action of the artificial viscosity $\mathcal Q$.

These equations are
solved using time-explicit Eulerian finite differencing.  The numerical
algorithm is that employed by the ZEUS code for hydrodynamics (Stone \&
Norman 1992a) and MHD (Stone \& Norman 1992b; Hawley \& Stone 1995).
Time and length units are established by setting $GM = R_g = 1$.   To
ensure a substantial reservoir of matter and to minimize the influence
of the outer boundary, the radial grid runs from $R=1.5$ to $R=61.5$
using 256 radial grid zones.  These grid zones are either evenly
spaced, or graded so as to concentrate resolution in the inner
regions.  The azimuthal angle $\phi$ runs from 0 to some fraction of $2\pi$
depending upon the simulation.  In the vertical direction $z$ runs over
a somewhat arbitrary periodic length. Periodic boundary conditions are
employed in $\phi$ and $z$.  The radial boundary conditions are simple
zero-gradient outflow conditions; no flow into the computational domain
is permitted.  The radial magnetic field boundary condition is set by
requiring the transverse components of the field to be zero outside the
computational domain, while the perpendicular component satisfies the
divergence-free constraint.

In the pseudo-Newtonian potential the angular momentum of a circular
orbit (here referred to as the Keplerian angular momentum) is
\begin{equation}\label{lkep}
\ell_{kep} = (GMR)^{1/2} {R\over R-R_g} = \Omega R^2.
\end{equation}
With $GM = R_g = 1$ the innermost marginally stable circular orbit
$r_{ms}$ is located at $R=3$; at this point $\Omega = 0.29$, and the
orbital period is $P_{\rm orb} = 2\pi/\Omega = 21.8$.  
At the grid outer boundary
the orbital period is almost 3000.  The advantage of the large radius
of the outer boundary is that many orbits can elapse in the inner
regions of the disk before the outer boundary conditions become
important to the simulation.  This offers the possibility of
establishing an accretion flow at the inner edge of
the disk independent of the outer boundary condition (although still
dependent of the initial conditions chosen for bulk of the disk).

In reporting results from the simulations many of the 
values will be averaged over both height $z$ and angle
$\phi$.  For example, the averaged mass density is
\begin{equation}\label{sigma}
\langle \rho\rangle = {\int \rho R d\phi dz \over \int R d\phi dz}.
\end{equation}
The net local mass flux is 
\begin{equation}\label{mdot}
\langle \dot M\rangle = \int -\rho v_R R d\phi dz,
\end{equation}
here defined to be positive for net accretion (inflow).  Similar
averages are constructed for pressure, magnetic energies, stresses and
angular momentum.  These averaged values are computed at specific time
intervals.  Where desired one can further average these quantities over
time.

A drawback to reporting space and time averaged values is that they
blur a very real and important property of the flow: it is highly
variable.  One way to measure the 
azimuthal fluctuation level at a given time is with the quantity
\begin{equation}\label{fluctuation}
{\delta \rho \over \rho}\left(R\right) =
{1 \over \langle \rho\rangle}  \left\{{1 \over 2\pi} \int \, d\phi \,
  \left[\rho - \langle \rho \rangle \right]^2 \right\}^{1/2} ,
\end{equation}
where ${\langle \rho \rangle}$ is defined in (\ref{sigma}).  

The azimuthal structure in the disk is examined with
a fourier transform of a vertically averaged
value such as  density, e.g.,  
\begin{equation}\label{fft}
\tilde \rho (R,m) = {1\over 2\pi} \int_{0}^{2\pi}  
\int_{-z_{max}}^{z_{max}} \rho(R,\phi,z)
e^{i m\phi} dz d\phi .
\end{equation}
The power, $|\tilde\rho |^2$, is further averaged over
radius within the interior of the active region in the disk.
Similarly, the time-dependent behavior of a disk is examined with a
fourier transform over time of azimuthally and vertically averaged
values.

Table 1 lists the simulations carried out as part of this study.
The models are identified by labels:  CK stands for cylindrical
Keplerian, HK for hydrodynamic Keplerian, and NK for Newtonian
Keplerian, two models that use the standard $1/R$ Newtonian potential
rather than the pseudo-Newtonian potential.  The table also lists the
computational domain, grid resolution, equation of state, initial
field topology, and end time in code units.

\section{Simulation Results}

\subsection{Disk with initial toroidal field}

The first Keplerian disk simulation, CK5, considers a disk with an
inner edge at $R=10$, set back from $r_{ms}$.  This
will investigate disk accretion removed from the dynamical forces
associated with the pseudo-Newtonian potential's innermost circular
orbit.  The simulation follows the global development of the MRI,
turbulent angular momentum transport, and the resulting gradual
accretion of matter and inward drift of the disk edge.

The initial conditions consist of an isothermal gas at constant
density, $\rho = 1$,  and a sound speed
$c_s = 0.03054$ which gives a Mach number ${\mathcal M}
= v_\phi/c_s = 11.5$ at $R=10$.  With the decrease in velocity
$v_\phi$ the Mach number declines to 4.3 by the
outer radial boundary.  The azimuthal angle $\phi$ runs from 0 to
$\pi/2$ in 58 grid zones, and in the vertical direction $z$ runs from 0
to 2 in 24 grid zones.  The vertical scale height at $R=10$ corresponds
to 0.87.  Of course in the cylindrical disk approximation the absence
of a vertical gravitational force means that the scale height has no
hydrodynamical significance in $z$; it is merely a convenient
expression of the temperature of the disk in terms of the ratio of the
sound speed to orbital frequency.

With a magnetic field, the vertical domain size retains significance
even in the cylindrical disk limit in relation to the most unstable
wavelength of the MRI.  The magnetic field strength determines this
wavelength, and the initial pressure determines the value of $P/P_{mag}
\equiv \beta$ to which this field strength corresponds.  For CK5 the
initial magnetic field is purely toroidal with $\beta = 4$.  For this
strength of field the ratio of the Alfv\'en speed to the orbital
frequency, $v_A/\Omega$, ranges from 0.6 at the inner disk boundary to
10.2 at $R=60$.  The critical azimuthal wavenumber for the MRI,
$m=R\Omega/v_A$, ranges from 16 at the inner disk to 6 at the outer
boundary.  Since the azimuthal direction spans $\pi/2$ with 58 grid
zones, only wavenumbers $4 \le m \le 116$ will be realizable.  With the
chosen magnetic field strength, however, the unstable azimuthal 
wavenumbers of the MRI should be well resolved throughout the disk.

The simulation is run out to time $t=3972$ which is 22 orbits at the
disk's initial inner boundary, and 180 orbits at $r_{ms}$.  As the
simulation proceeds the MRI develops, first at the inner disk edge and
spreading outward through the disk as time advances.  Small wavelength
perturbations (large $m$) grow most rapidly, but the largest azimuthal
scales dominate in the end.  The exponential growth of the poloidal
field at a given radius ends after approximately five orbits, as
measured locally.  The Maxwell stress in the resulting MHD turbulence
transports angular momentum and the disk evolves.  Figure 1 shows a
spacetime $(R,t)$ plot of the Maxwell stress, $M^{R\phi}$, in terms of
a Shakura \& Sunyaev (1973) $\alpha$ using the initial pressure as a
normalization, $\alpha = M^{R\phi}/\rho c_s^2$.  This plot illustrates
the rapid local buildup of magnetic stress to significant levels, as
well as the large fluctuations in time and space.

\begin{figure}
\centerline{\psfig{file=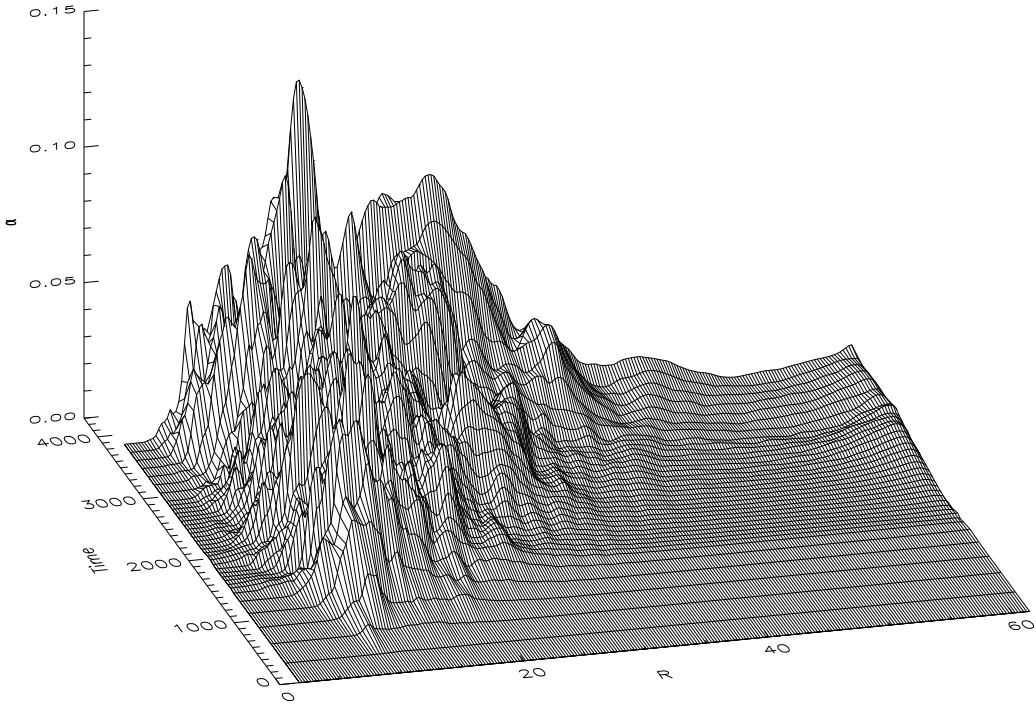,width=6.0in,angle=90}}
\caption{Spacetime diagram of the Maxwell stress in run CK5 
normalized by the
initial pressure $\alpha=M^{R\phi}/\rho c_s^2$.  The stress develops 
first at $R=10$ and then spreads out through the disk as the MRI 
grows at a rate proportional to the local orbital frequency.  
At the end of the simulation the stress rises rapidly from near
zero at $R=5$, reaches the highest levels between $R=15$ and 30,
then declines rapidly with radius.  Within the disk
the stress varies strongly in both time and space. }
\end{figure}

Overall, the disk never attains a steady state; its inner edge slowly
moves inward and fluid in the outer disk moves outward.  Through the
course of the simulation the disk inner edge moves from $R=10$ down to
$R=5$.  This is illustrated in Figure 2 which shows $\langle\rho\rangle$
at several times during the run.  At the end time, the disk
has not quite reached $r_{ms}$ and there is no significant mass flux
through the marginally stable orbit and into the central hole.
Magnetic field has been carried inward along with the mass.  The
regions with the strongest fields lie between the peaks in the density
distribution.  The average $\beta$ value varies between 8 at the
density peaks and 2 in between.

\begin{figure}
\centerline{\psfig{file=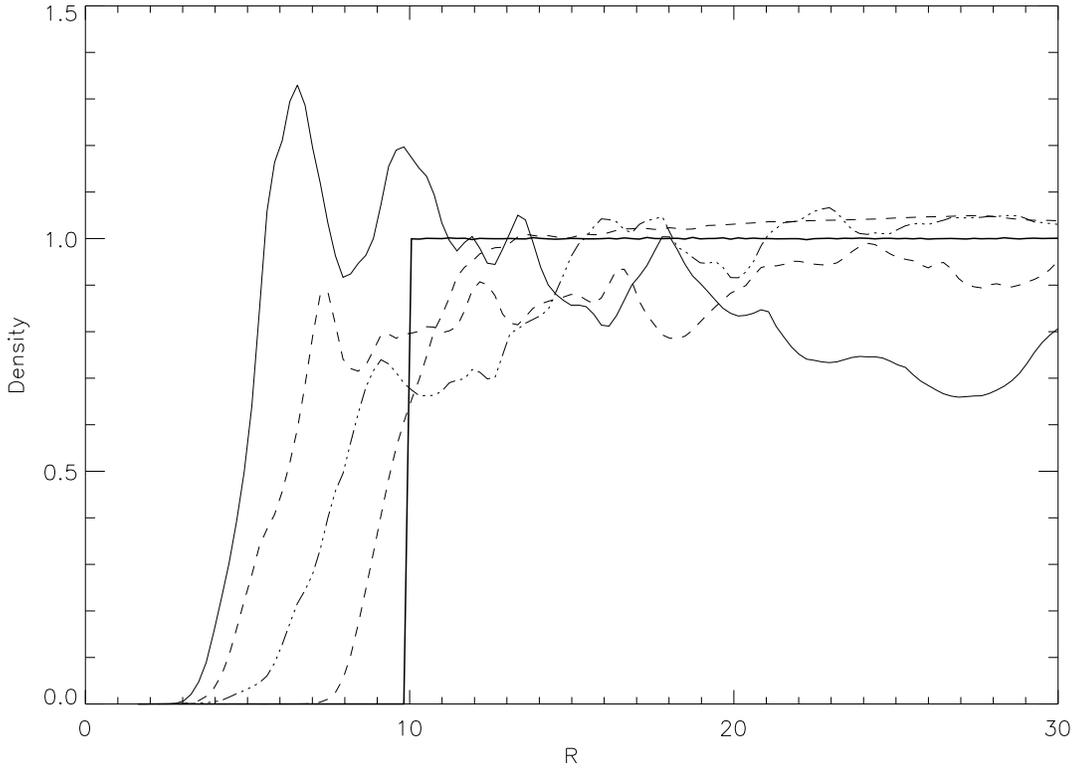,width=6.0in,angle=90}}
\caption{Radial profile of density $\langle\rho\rangle$ at 
times $t=0$ (solid bold line), 890 (short dashed line), 1835
(dash-dotted line),
2734 (thick dashed line), 3848 (solid thin line) 
in simulation CK5.  These times correspond to 0, 4.9, 10.2,
15.2, and 21.5 orbits at $R=10$.  Angular momentum transport produced by
MHD turbulence causes the inner edge of the disk to drift gradually
inward.
}
\end{figure}

Averaged properties of the accretion flow within the disk can be
derived from a region between $R=13$ and 22 which comes into an
approximate time-averaged steady state after $t=2800$.  The
time-averaged accretion rate within this region is roughly constant
with radius, with $\dot M = 0.075$, and $v_R/v_\phi = -0.006$.  The
instantaneous fluctuations in radial velocity, $\delta v_R/v_\phi$
typically range between $0.01$ and $-0.02$.  Within
the turbulent disk the toroidal field energy dominates over that of the
poloidal field, with the ratio $B_\phi^2/B_R^2 = 13$.  The radial field
energy is, in turn, a factor of 6 greater than the that of the vertical
field.  The toroidal field energy is essentially unchanged in the mean
from its initial value, although there are spatial variations in
strength by a factor of two within the disk.  At the end of the run the
region of significant Maxwell stress has moved out to around $R=40$
with the largest values lying between $R=7$ and 30 (Fig. 1).  The
time-averaged Maxwell stress over this range is about $5.5\times
10^{-5}$, corresponding to an average value of $\alpha = 0.07$.  The
relatively low level of the radial field compared to the toroidal is
reflected in the value of $\alpha_{mag}$, the ratio of the Maxwell
stress to the total {\it magnetic} pressure ($\alpha = \alpha_{mag}/\beta$).
In the inner disk this value ranges between 0.2 and 0.3.  The toroidal
field simulation of ARC obtained an $\alpha$ slightly below 0.01, and
an $\alpha_{mag} \sim 0.3$.  (Note that there are differences in  
$\alpha$ terminology between this paper and ARC.)

The averaged magnetic properties of the turbulence are consistent with
the results of toroidal field local shearing box simulations (cf. Table
3 of HGB95).  For those simulations $\alpha$ was typically
a few times 0.01, and $\alpha_{mag} \sim 0.4$ for simulations that
began with weak toroidal fields and $\sim 0.2$ for those that 
began, like this simulation, with stronger toroidal fields.

\subsection{Disk with initial vertical field}

The two general classes of models considered in the local shearing box
limit by HGB95 were initial vertical fields and initial toroidal
fields.  In the next simulation, CK6, we complete the extension of
these local models from the shearing box to the cylindrical disk by
simulating the same disk as CK5, but now with an initial vertical
magnetic field.

Simulation CK6 is a Keplerian disk consisting of an
isothermal gas at constant density $\rho = 1$ from $R=10$ to the outer
boundary, and a constant sound speed $c_s = 0.03054$.  The radial grid
resolution is the same as in CK5, but a few additional zones are added
in the $\phi$ and $z$ direction;  $\phi$ runs from 0 to $\pi/2$ over 64
grid zones, and $z$ runs from 0 to 2 in 32 grid zones.  In CK6  the
strength of the initial vertical magnetic field is set by  requiring
$2\pi v_A/\Omega=0.5$ at every radius.  This is nearly the fastest
growing wavelength of the linear MRI; there will be four such
wavelengths across the $z$ domain.  The initial magnetic
field decreases with radius (proportional to $\Omega$), with values of
$\beta$ ranging from 240 at the inner edge of the disk to 66,000 at the
outer grid boundary.  Although the most unstable vertical wavelength is
the same throughout the disk, the local growth rate of the MRI is
proportional to the local orbital frequency, so the magnetic
instability grows more slowly with increasing radius.

The simulation is run out to time $t=4097$ which is 22.9 orbits at the
disk's initial inner boundary ($R= 10$), 188 orbits at $r_{ms}$, and
only 1.37 orbits at the grid outer boundary.  As in CK5, MHD turbulence
develops first in the inner disk and spreads outward.  In CK6 the
instability grows more rapidly, and although the field is much weaker
initially, it still grows to about the same level as in CK5.

In CK6 accretion is driven through the marginally stable orbit at early
time.  This is due, in part, to a unique property of the vertical field
instability.  Large fluctuations and impulsive accretion are created by
the ``channel solution'' of the nonlinear vertical field instability
(Hawley \& Balbus 1992; HGB95).  The fastest growing modes have finite
vertical wavenumber $k_z$ and vanishing radial and azimuthal
wavenumbers, $k_R$ and $m$.  This leads to coherent radial flow upon
nonlinear saturation.  Although the development of full turbulence
eventually limits the coherence of this radial channel flow, early on
in the simulation at the inner edge of the disk radial filaments of gas
are able to move rapidly inward and accrete.

As the simulation proceeds, an interesting structure emerges in the
inner part of the disk.  By $t=2000$ gas has piled up in a density
maximum located at $R=7.5$ and in a smaller maximum at $R=4$.  Between
these rings are local minima at $R=5$ and $R=10$.  These can be seen in
Figure 3,  a spacetime diagram of the vertically- and
azimuthally-averaged density.  For clarity, only the inner half of grid
is shown and only up to $t=2000$.  A local maximum in density
corresponds to a local minimum in the Maxwell stress and the Alfv\'en
speed and vice versa.  As time proceeds, the inner density maximum
accretes into the hole, while the other slowly drifts inward, reaching
$R=4.2$ by $t=4096$.  By this time another density gap is opening up
near $R=17$.

\begin{figure}
\centerline{\psfig{file=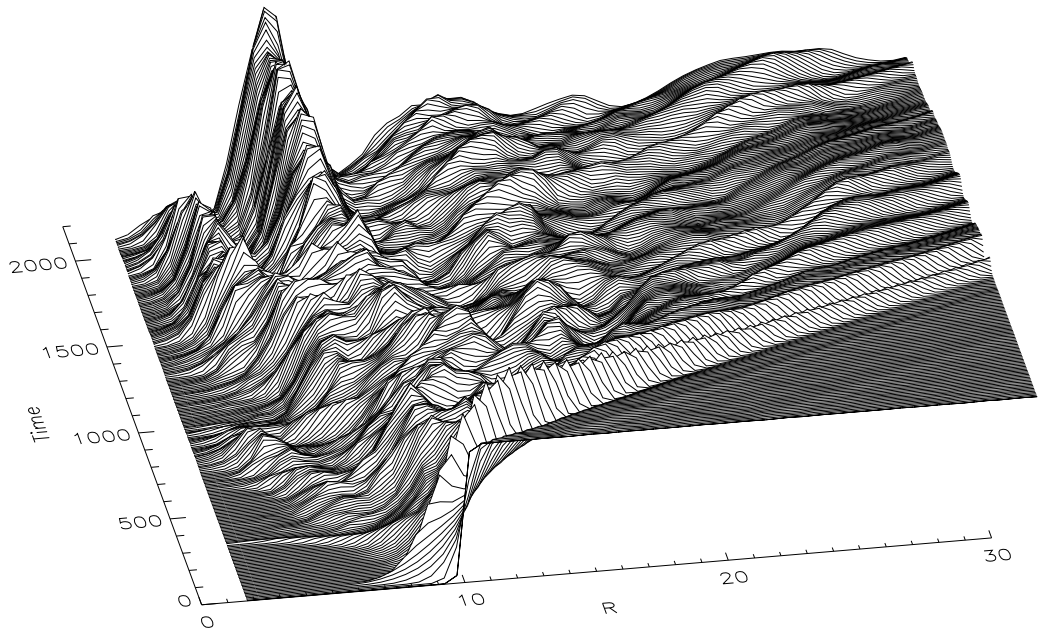,width=6.0in,angle=90}}
\caption{$(R,t)$ spacetime diagram of the vertically and azimuthally
averaged density $\langle\rho\rangle$ in simulation
CK6 for the inner half of the radial grid from $t=0$ to 2000.  The disk
is not initially in hydrostatic equilibrium at the inner edge,
resulting in an outward traveling wave after $t=0$.  The turbulence
itself subsequently generates magnetoacoustic waves.  Of
particular interest are the local density maxima and the gaps
surrounding them which develop after $t=1000$ at $R=5$ and $R=10$. 
}
\end{figure}

The specific angular momentum, density, and Maxwell stress at the end
time are shown in Figure 4.  The Maxwell stress, and hence the angular
momentum transport, is larger in regions of low density.  The radial
slope of the specific angular momentum $\ell$ becomes steeper in the
regions of maximum Maxwell stress, and flatter in the density maxima.
Essentially, the disk is forming a local pressure-supported slender
torus with non-Keplerian angular momentum distribution to balance the
internal pressure forces.  Gas moves rapidly out of the region between
the rings, but is blocked from further accretion by the torus itself.
Thus there is a tendency for material to concentrate into rings.

The presence of dense rings and empty gaps is a description of the
behavior of a ``viscous-type'' instability (Lightman \& Eardley 1974)
where the stress is a decreasing function of the density.  Here the
cause of this behavior is different from that considered originally by
Lightman and Eardley (1974).  It is not a property of the Maxwell
stress {\it per se} as much as it is a consequence of the properties of
the MRI.  Lower density corresponds to larger Alfv\'en speeds which, in
turn, produce faster growth rates at larger wavelengths of the MRI.
Longer wavelengths are more efficacious in transporting larger amounts
of angular momentum.

\begin{figure}
\centerline{\psfig{file=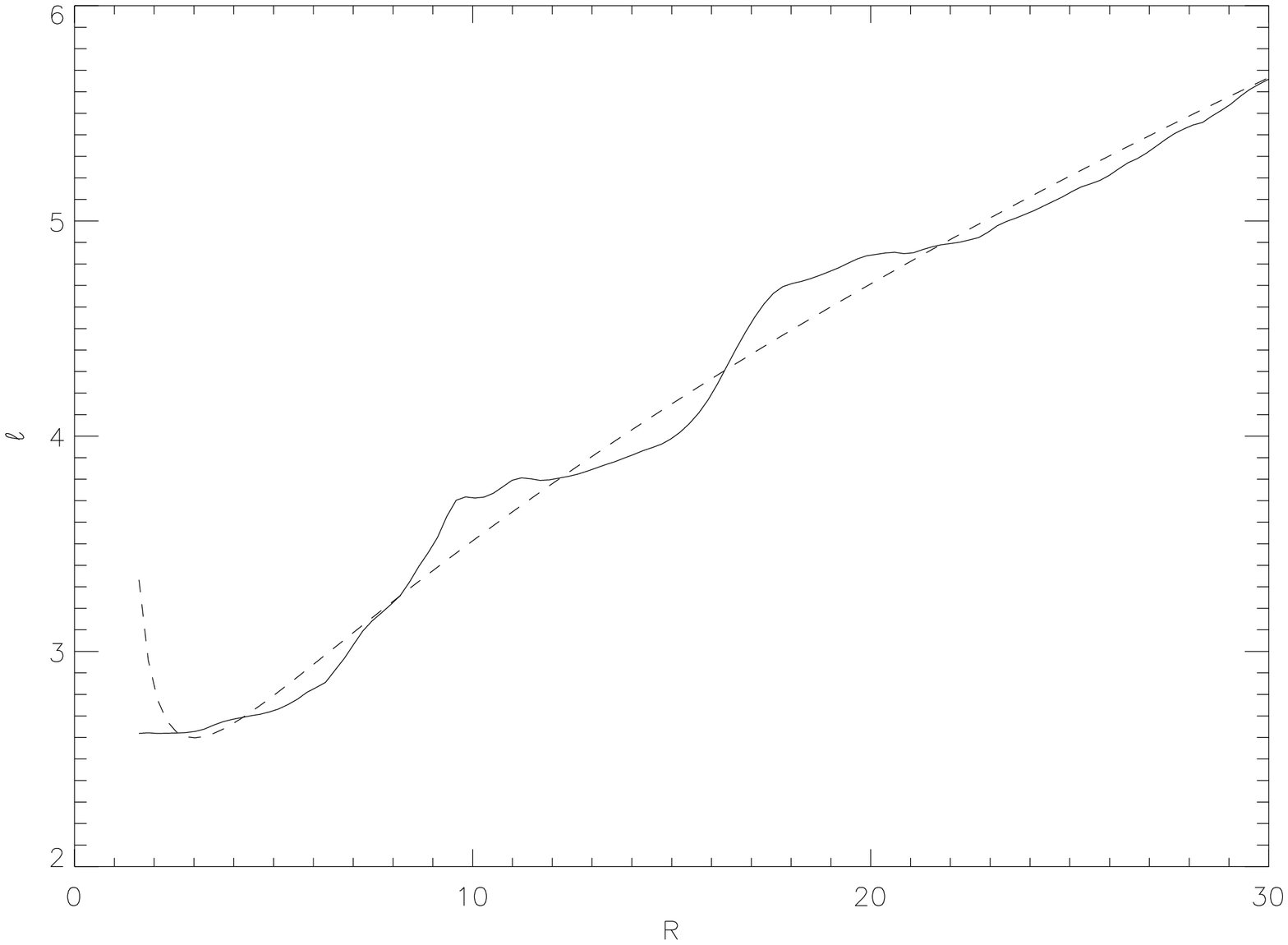,width=3.5in,angle=90}}
\centerline{\psfig{file=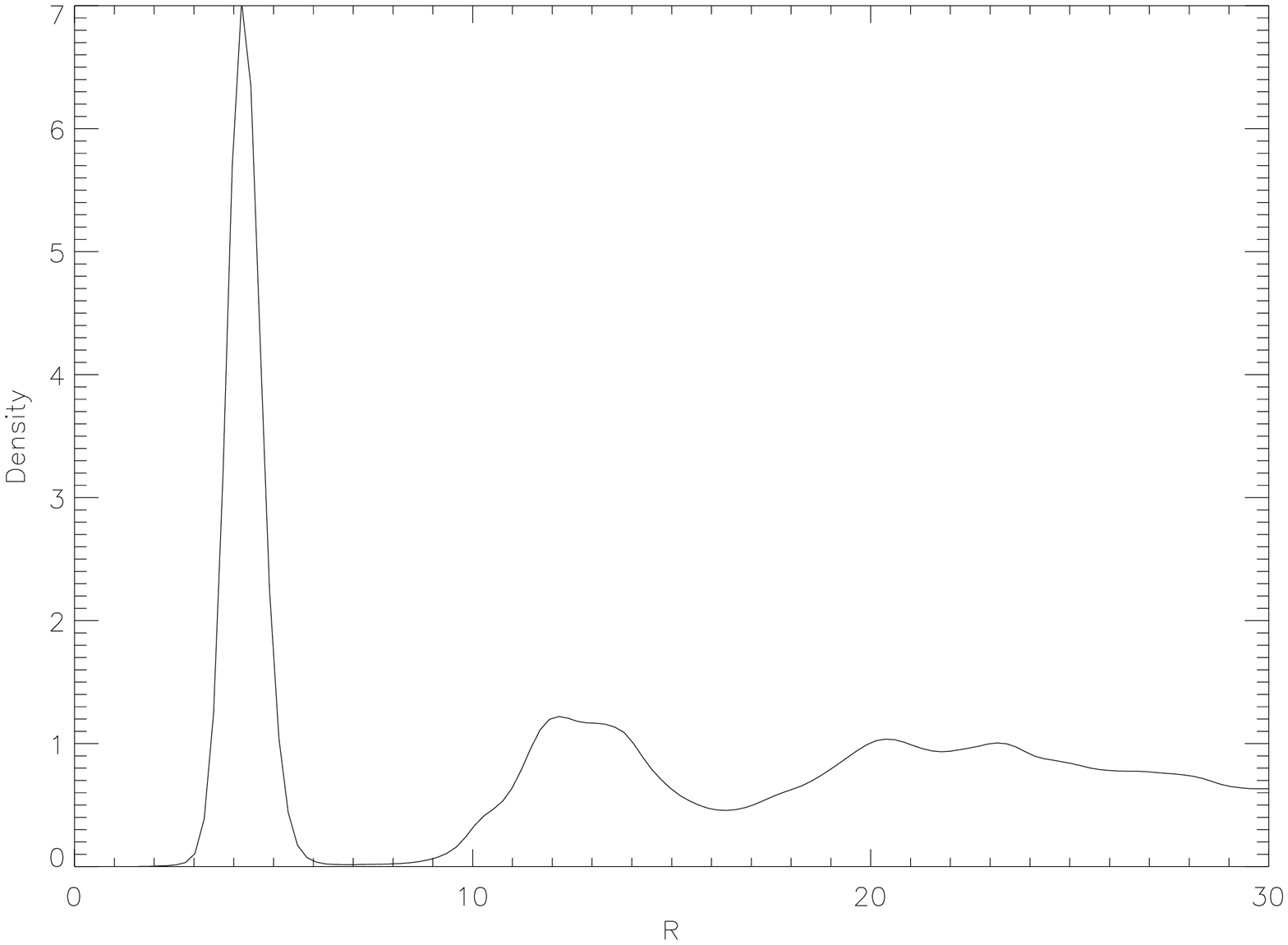,width=3.5in,angle=90}}
\centerline{\psfig{file=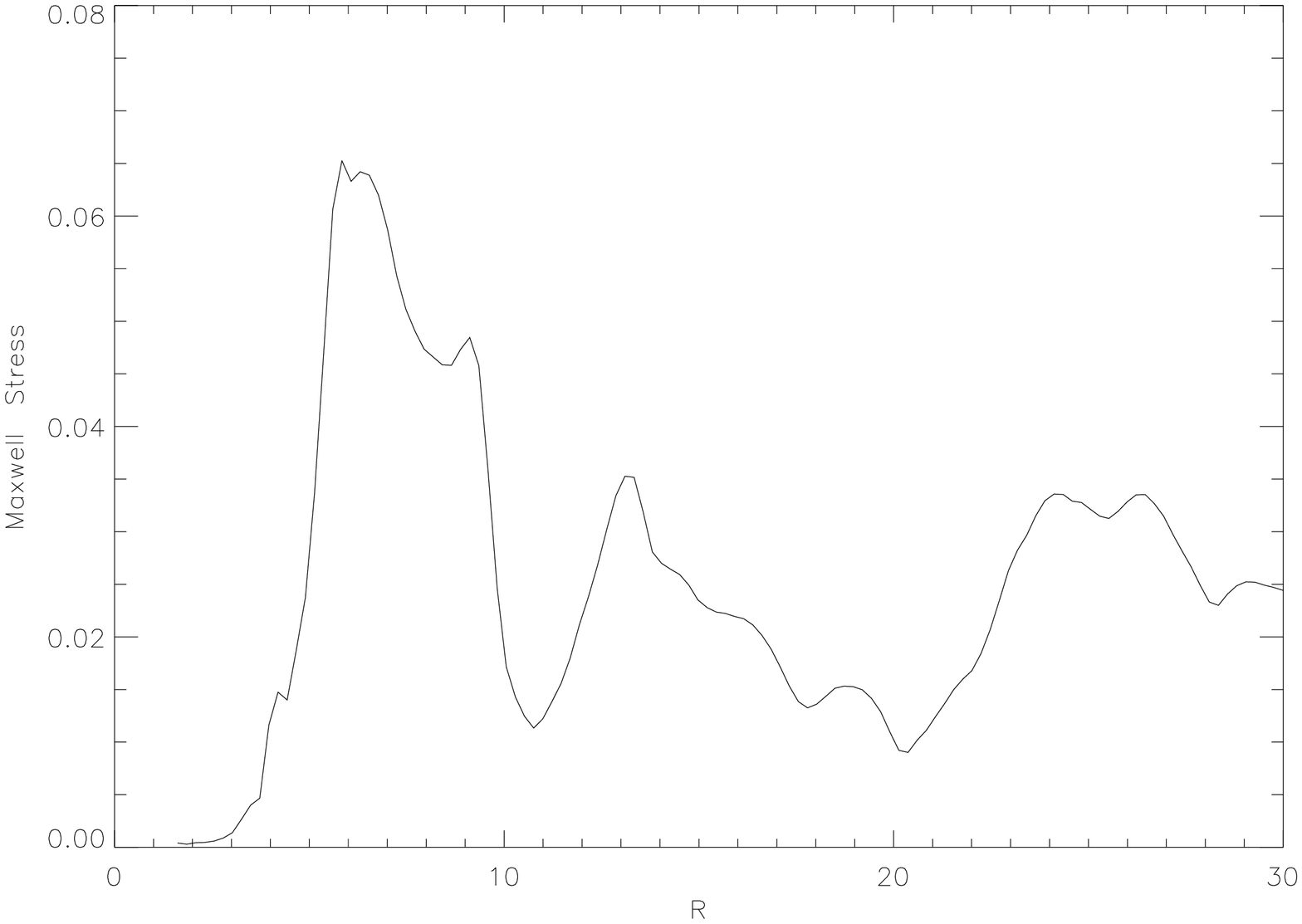,width=3.5in,angle=90}}
\figcaption{Vertically- and azimuthally-averaged 
values at $t=4096$ in simulation CK6.
(a) Specific angular momentum $\langle \ell\rangle$ overlaid on the Keplerian
angular momentum (dashed line), (b) density $\langle\rho\rangle$,
and (c) Maxwell stress normalized by the {\it initial} gas pressure,
$M^{R\phi}/\rho c_s^2$.
}
\end{figure}

A spacetime plot of the accretion rate shows evidence for significant
time variability and periodicity.  The variability is manifest as
pressure waves that propagate throughout the disk at the sound speed.
A fourier transform of the spacetime data reveals peaks in the power
spectrum which originate at various radii in the disk.  The peaks occur
where the pressure waves are generated at frequencies close to the
local orbital frequency.  Initially the largest peak in the spectral
energy is found close to $R=13$;  this is also where the peak Maxwell
stress is generated.  At late time the most prominent frequency observed
corresponds to the orbital frequency at the location of the dense
inner ring.  This, in turn, drives a periodic accretion flux
into the central hole.

The substantial spatial inhomogeneity of the disk at late time makes it
difficult to assign overall ``steady state'' values to quantities.  The
azimuthal density fluctuations, characterized by (\ref{fluctuation})
are as large as 0.3 in the region of maximum Maxwell stress, and are
between 0.1 and 0.2 throughout the bulk of the radial extent of the
disk.  This level is comparable to that seen in the toroidal field
simulation CK5.  What sets CK6 apart are the large variations with
radius.  

The magnetic field has undergone substantial amplification.  At the end
of the run $\beta$ is about 70 inside the dense ring at $R=4$, and
falls to $0.2$ in the low density region at $R=7$.  From $R=10$ to 40,
$\beta$ has a minimum of $14$, with a local maximum $\sim 70$ at
$r=20$.  Beyond $R=30$ $\beta$ slowly rises with radius to around 100
at the outer boundary.  As stated above, the Maxwell stress is largest
between the dense rings.  The Shakura-Sunyaev $\alpha$ is as low as
0.002 in the inner ring, rising to $\sim 4$ at $R=7$ before dropping
back to values ranging from $0.01$ to 0.05 between $R=10$ and 40.  As
in run CK5 the average total toroidal field energy dominates over that
of the poloidal field, although here by a lesser amount,  with
$B_\phi^2/B_R^2 = 6.3$.  The radial field energy is a factor of 6
greater than the vertical field energy.  The ratio of the Maxwell
stress to the magnetic pressure, $\alpha_{mag}$, varies with radius,
but has an average of 0.5.  This is larger than seen in CK5, and is the
typical value  for local shearing box simulations with a net vertical
field.

\subsection{Disk near the marginally-stable orbit}

The recent work of HK and ARC focused on the stress at and interior to
the marginally stable orbit.  Having considered disks with inner
boundaries located outside of $r_{ms}$, we now turn to a model, CK7, a
disk with an initial inner edge near $r_{ms}$ that produces substantial
accretion into the central hole.  It uses an adiabatic equation of
state with $\Gamma = 5/3$.  The computational grid extends from $R=1.5$
to $R=61.5$, over 0.8 in $z$, and $2\pi$ in $\phi$.  There are 32
equally spaced zones in $z$, 256 equally spaced zones in $\phi$, and
256 zones in $R$.  Eighty of the radial zones are equally spaced
between $R=1.5$ and 10, while the remainder are logarithmically
stretched between 10 and 61.5.  As before, the large radial extent of
this disk provides a reservoir of mass which will allow the inner
region of the disk to evolve for many orbits without influence from the
outer grid boundary.

For this simulation the initial state was relaxed to a true
hydrodynamic equilibrium at the inner disk boundary.  A constant
density gas with $\rho = 1$ is first placed on the grid from $R=4$ to
the outer boundary.  The initial sound speed is chosen to be $c_s =
0.086$ throughout the disk, corresponding to an isothermal scale height
of 0.4 at $R=4$ ($c_s^2/\Gamma = 0.01 v_\phi^2$).  This is hotter than
the standard model of ARC ($c_s = 0.069$), but cooler than the thick
disk model of HK.  This initial condition is evolved in one dimension
to an equilibrium solution.  The resulting disk has $\rho = 1$ down to
about $R=7$, inside of which the density and pressure decline smoothly
to zero at $R=3$.  Similarly, the angular momentum is Keplerian
throughout most of the disk, but becomes slightly super-Keplerian as
$R=3$ is approached.  The maximum increase over the Keplerian value is
3\%.

A vertical magnetic field is placed onto this hydrodynamic
equilibrium.  As in run CK6 the field strength is set so that the
Alfv\'en speed $v_A = \Omega/k_z$ for a wavenumber $k_z$ corresponding
to a wavelength of one quarter the vertical grid size.  This is equal
to $\beta = 276$ at $R=4$ with $\beta$ increasing to $10^6$ by $R=60$.

This disk is evolved for 270,000 timesteps out to a time $t=2575$.
This corresponds to 118 orbits at $r_{ms}$.  Accretion through the
inner boundary begins with a strong pulse at $t\sim 100$ as the
magnetic instability in the inner part of the disk reaches nonlinear
saturation.  As the evolution proceeds, the location where magnetic
instability is undergoing its initial linear growth moves out through
the disk.  The region that is in a quasi-steady state similarly moves
out through the disk, albeit at a slower rate.  The resulting disk can
be described both in terms of instantaneous properties, which emphasize
the turbulence and the fluctuations, or in terms of time- and
space-averaged quantities which provide some measure of what the
``steady state'' properties of the disk are like.

Figure 5 shows the surface density distribution at the end time of the
simulation.  The most prominent features are the tightly wrapped
trailing spiral waves.  The spiral waves develop in the inner part of
the disk along with the turbulence.  A spacetime diagram shows that
these waves propagate out through the disk at the sound speed.  The
waves are produced quasi-periodically throughout the evolution.   A
fourier transform of the accretion rate shows that the peak spectral
energies correspond to the orbital period at $R=6.7$ and 5.6 in the
first half of the simulation, and larger radii at later times as the
turbulence grows at those locations.  The amplitude of the density
fluctuations through these waves is measured by equation
(\ref{fluctuation}).  Outside of $R=20$ $\delta\rho/\rho \simeq 0.2$.
Inside of this point, the fluctuation amplitude rises, reaching a value
of 1.3 at $r_{ms}$.  The accretion flow through $r_{ms}$ tends to be
highly nonaxisymmetric.

\begin{figure}
\centerline{\psfig{file=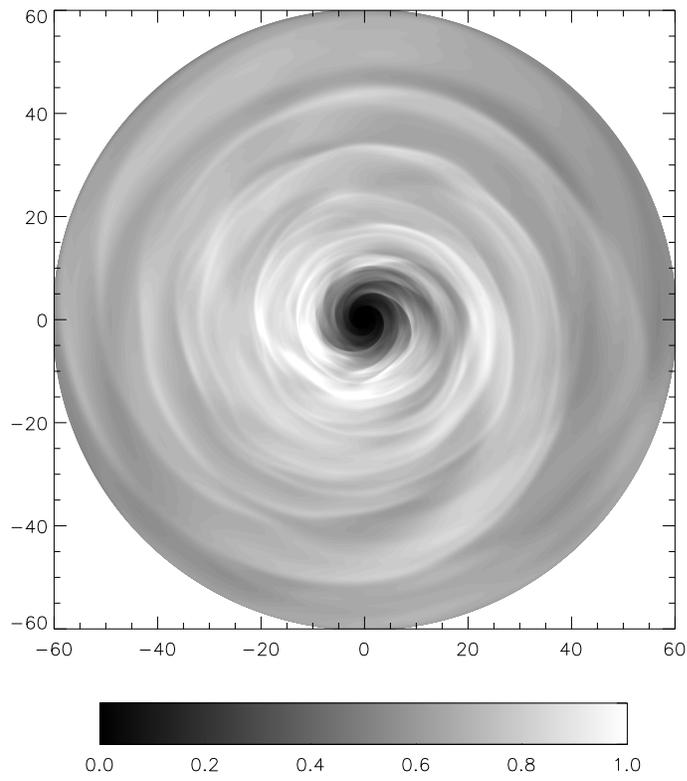,width=6.0in,angle=90}}
\figcaption{Vertically-averaged gas density in run CK7 at the end 
time $t=2575$.  The grey-scale is linear in density and runs
from 0 to 1.
}
\end{figure}

Of particular interest for this simulation is the magnitude of the
stress at the marginally stable orbit, and its effect on the specific
angular momentum $\ell$.  Throughout most of the disk the averaged
$\ell$ tracks the circular orbit value, $\ell_{kep}$, with a one to two
percent excess over this value inside $R=10$.  Figure 6 shows the time
evolution of both the Maxwell stress at $r_{ms}$ in terms of the
Shakura-Sunyaev $\alpha$ value defined with the local pressure, and the
specific angular momentum at $r_{ms}$ and close to the inner boundary
at $R=1.66$.  The first thing to notice is that the Maxwell stress at
$r_{ms}$ is nonzero and time-variable, as are the values of $\ell$ at
$r_{ms}$ and the inner boundary.  There is, however, always a net
change in $\ell$ between $r_{ms}$ and the inner boundary. At $r_{ms}$,
$\ell$ is always slightly greater than $\ell_{kep}\approx 2.6$; the
time-average excess in the second half of the simulation is 1.8\%.  And
although the slope $d\ell/dR$ decreases toward zero inside of
$r_{ms}$,  there is always an average decline in $\ell$ of 1.7\%
between $r_{ms}$ and the inner radial boundary.  This is smaller than
the 5\% drop reported by HK, but appears to be consistent with the
results of ARC.

\begin{figure}
\centerline{\psfig{file=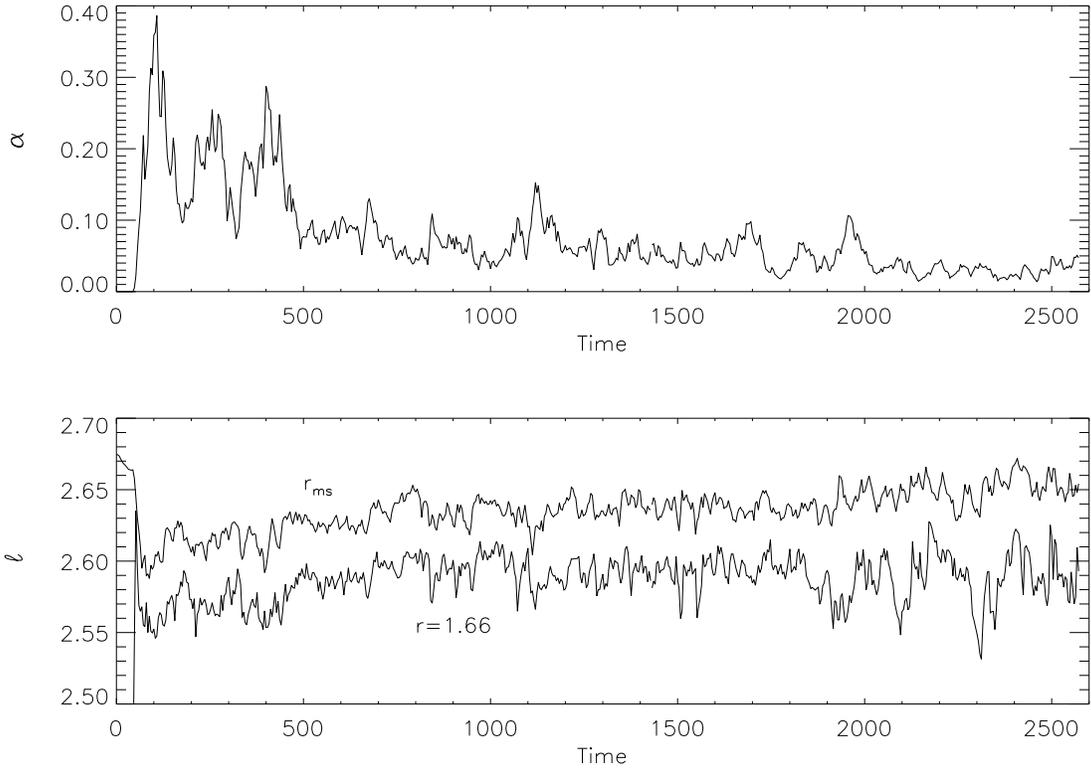,width=6.0in,angle=90}}
\figcaption{a) Maxwell stress $M^{R\phi}$ in CK7 at $R=r_{ms}$ as a
function of time in terms of $\alpha=M^{R\phi}/P$,
where $P$ is the local gas pressure.  (b) Time history of the specific
angular momentum at $R=r_{ms}$ and $R=1.66$.  Although $\ell$ varies
considerably with time, there is always a small net change between
$r_{ms}$ and the inner radial boundary. }
\end{figure}

Figure 7 shows the $\phi$- and $z$-averaged magnetic energies and the
Maxwell stress as a function of radius averaged over the last 500 units
of time.  The values are scaled by the initial pressure.  As always,
the toroidal field energy is dominant, exceeding the radial field
energy by a factor that declines with radius, from 20 at $R=4$ to
around 5 outside of $R=15$.  Inside $R=3$, the radial field becomes
more significant compared to the toroidal as the field is combed out by
the nearly radial infall.  The ratio of the toroidal to the vertical
field energy is $\sim 70$ from $R=10$ to 20, and rises rapidly both
inside and outside this region.  At $r_{ms}$ $\beta = 8.5$; this
increases with radius to $\beta \approx 100$ at $R=20$.  Note that the
gas pressure is substantially reduced from its initial value inside of
$R=10$.  The magnetic energy has grown by over 600 times from its
initial value.  Beyond $R=20$ $\beta$ rises more gradually to about 400
at the outer boundary.

\begin{figure}
\centerline{\psfig{file=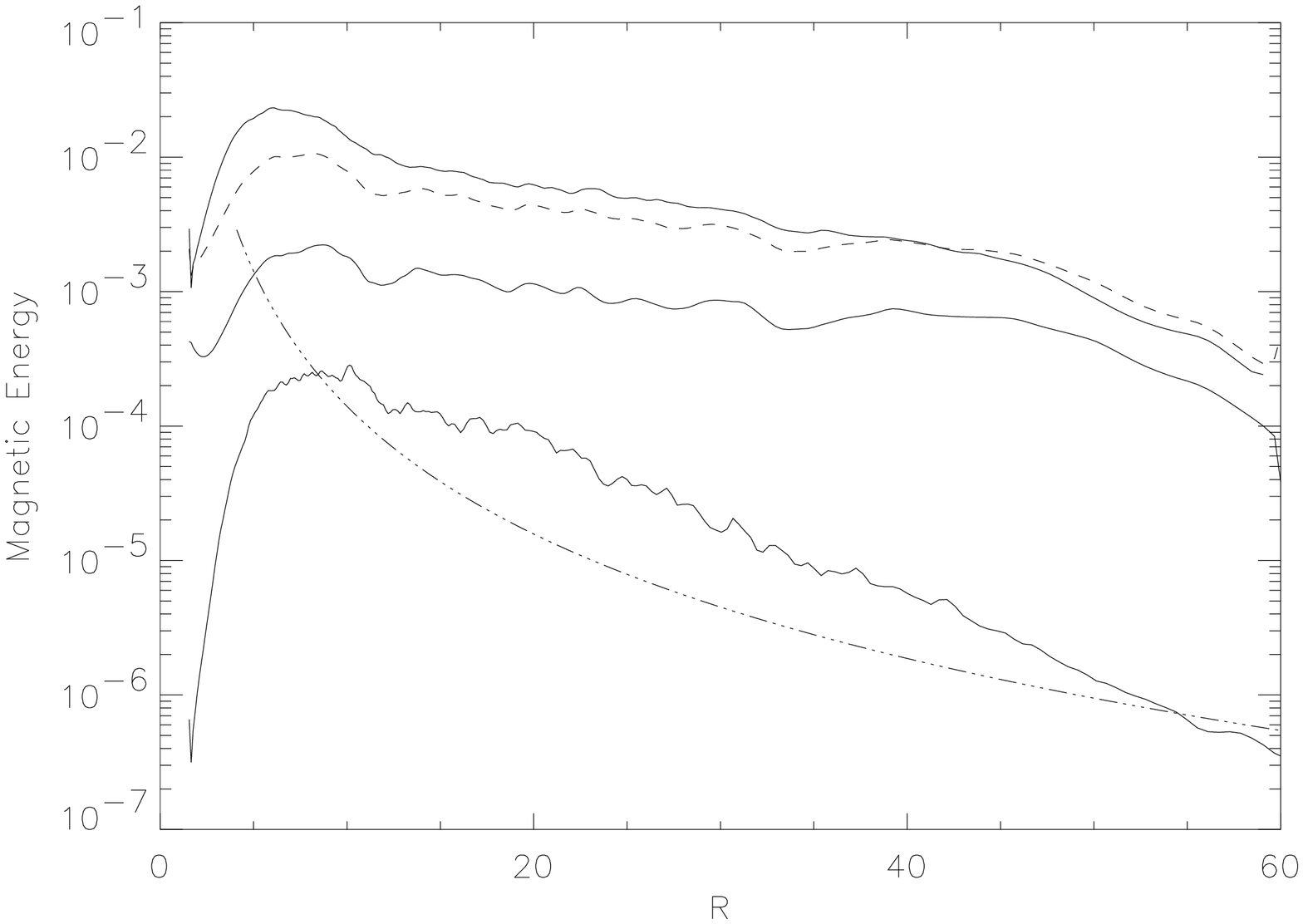,width=6.0in,angle=90}}
\figcaption{Magnetic energy averaged over $\phi$, $z$, and the last 
500 units of time in run CK7.  The top solid curve is the toroidal field,
the middle solid curve the radial field, the bottom solid curve the vertical
field energy.  The dashed line is the Maxwell stress, and the
dot-dash line is the initial vertical field energy.  All values are
normalized with the initial gas pressure ($P=0.0045$).
}
\end{figure}

The averaged mass accretion rate is relatively constant inside of
$R=15$ at late time.  Beyond $t=400$ the accretion rate into the
central  hole varies around a mean value of $0.072$.  The sense of
accretion reverses at $R=18$.  The averaged radial drift velocity has
the value $v_r/c_s \approx 0.008$ between $R=10$ and 15, and rises
rapidly to cross over $v_r/c_s=1$ at $R=2.6$; this is very
similar to the velocity plot (Fig. 4) of ARC.

\subsection{Influence of $\phi$ domain}

One of the potentially most useful approximations for three-dimensional
simulations is to reduce the azimuthal angular coverage to some integer
fraction of $2\pi$.  The advantage of this is obvious:  CK6 covered
$\pi/2$ in 64 grid zones, whereas CK7 required 256 grid zones to span
the full $2\pi$ with the same $\Delta \phi$.  Model CK7a investigates
the potential drawbacks of this reduction in computational domain size
by providing a direct comparison between a simulation spanning $\pi/2$
and one spanning the full $2\pi$ (CK7).  The grid size $\Delta \phi$ is
the same in both simulations.

The two simulations are very similar in their qualitative appearance,
although there are quantitative differences.  CK7a has about 10\% less
magnetic energy and magnetic stress on average.  On the other hand,
CK7a exhibits larger fluctuations in those quantities.  These
differences carry over into the accretion rate into the black hole.  In
CK7 $\dot M$ is 11\% larger on average, but CK7a has larger impulsive
spikes in accretion rate.

The larger fluctuation level seen in CK7a may be due, in part, to
existence of the channel solution for vertical fields.  Ideally the
channels have a finite $k_z$ and $k_R = m = 0$.  However, parasitic
instabilities (Goodman \& Xu 1994)  with nonzero $m$ and $k_R < k_z$
cause the breakup of the channel solution into smaller scale
turbulence.  Reducing the azimuthal extent of the computational domain
apparently makes it easier for the channel solutions to maintain 
some spatial coherence for slightly longer time.

Does the reduced domain influence the amount of stress at the
marginally stable orbit?  The cylindrical disk simulations of ARC
found a smaller decrease in specific angular momentum
between $r_{ms}$ and the inner boundary compared with the thick disk
simulation of HK.  Since ARC used an azimuthal
domain of only $\pi/6$ in angle it is possible that some of the
reduction in stress was due to the restricted angular domain.  In CK7a
the average $\alpha$ value at $r_{ms}$ after $t=1000$ is 0.053,
although $\alpha$ does briefly go as high as 0.15 at several points in
time.  The mean value of $\ell$ at $r_{ms}$ is 2.64 and the decrease
between that point and the inner boundary is 0.032, or 1.2\%.  This is
less than seen in CK7 (see Fig.~6), and is consistent with
the general reduction in average stress levels in the reduced domain
simulation.  Thus, although the smaller domain size reduces the
stress, this effect plays only a minor role in
the quantitative differences in the simulations of HK and ARC.

As with CK7, the MHD turbulence produces propagating pressure waves.
These are generated primarily in the inner portion of the disk at
frequencies corresponding to the orbital frequency (rather than the
periodicity frequency $\Omega/4$).  These waves are tightly wrapped,
low $m$ trailing spirals.  Figure 8a compares an angular power spectrum
of density $\rho$ for CK7 and CK7a at $t=1500$.  The spectrum is
averaged over radius from $R=5$ to 20.  Both simulations show
increasing power toward smaller azimuthal wavenumbers $m$.  This is
consistent with the visual appearance of the disk which in which low
$m$ spiral waves dominate.  The steepness of the 
slope of the power spectrum increases
with $m$, from $\sim -2$ at low $m$, up to $\sim -7$ for large $m$.
The two simulations have very similar power spectra;
CK7a is simply truncated by symmetry at $m=4$.

\begin{figure} 
\centerline{\psfig{file=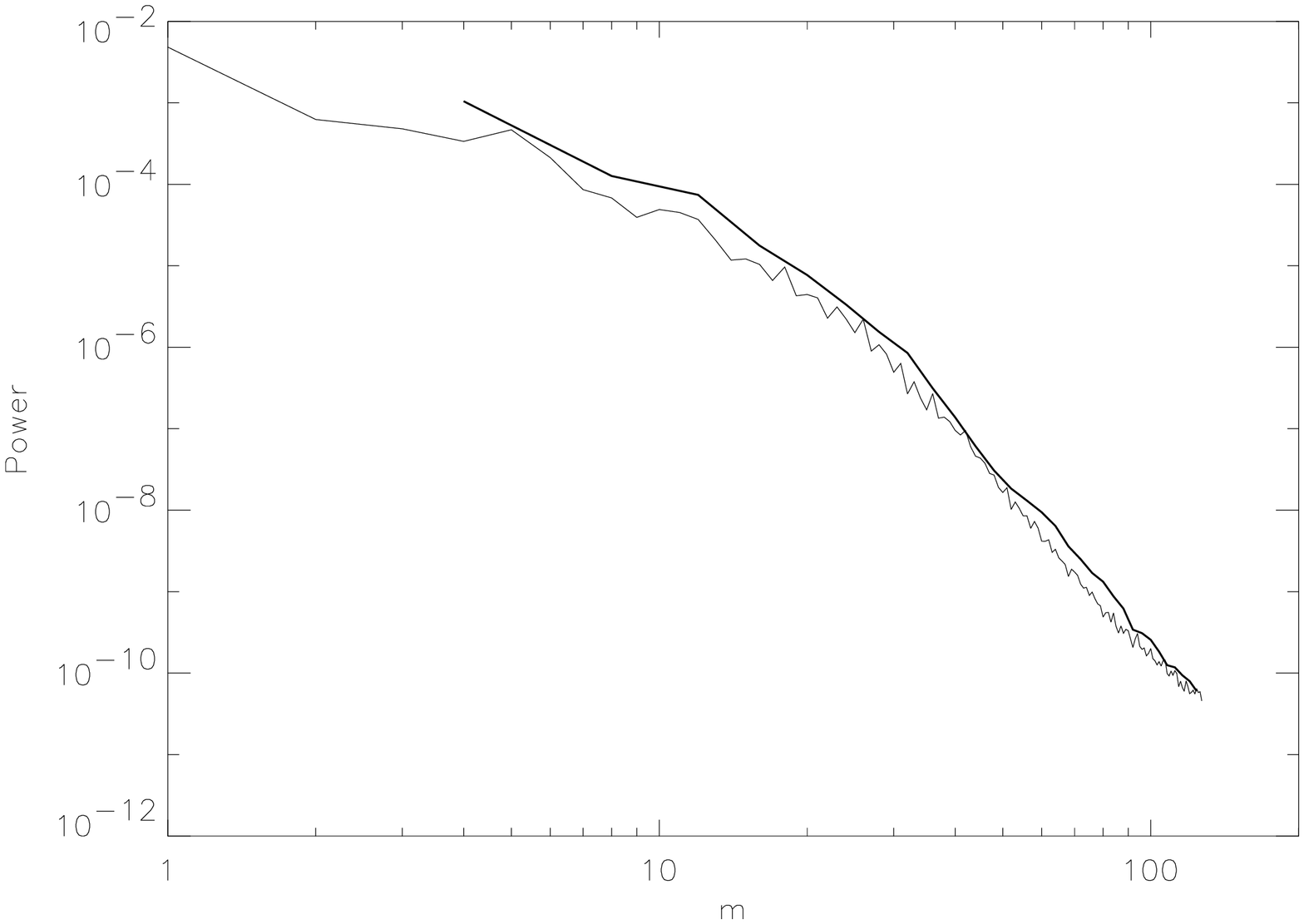,width=3.5in,angle=90}}
\centerline{\psfig{file=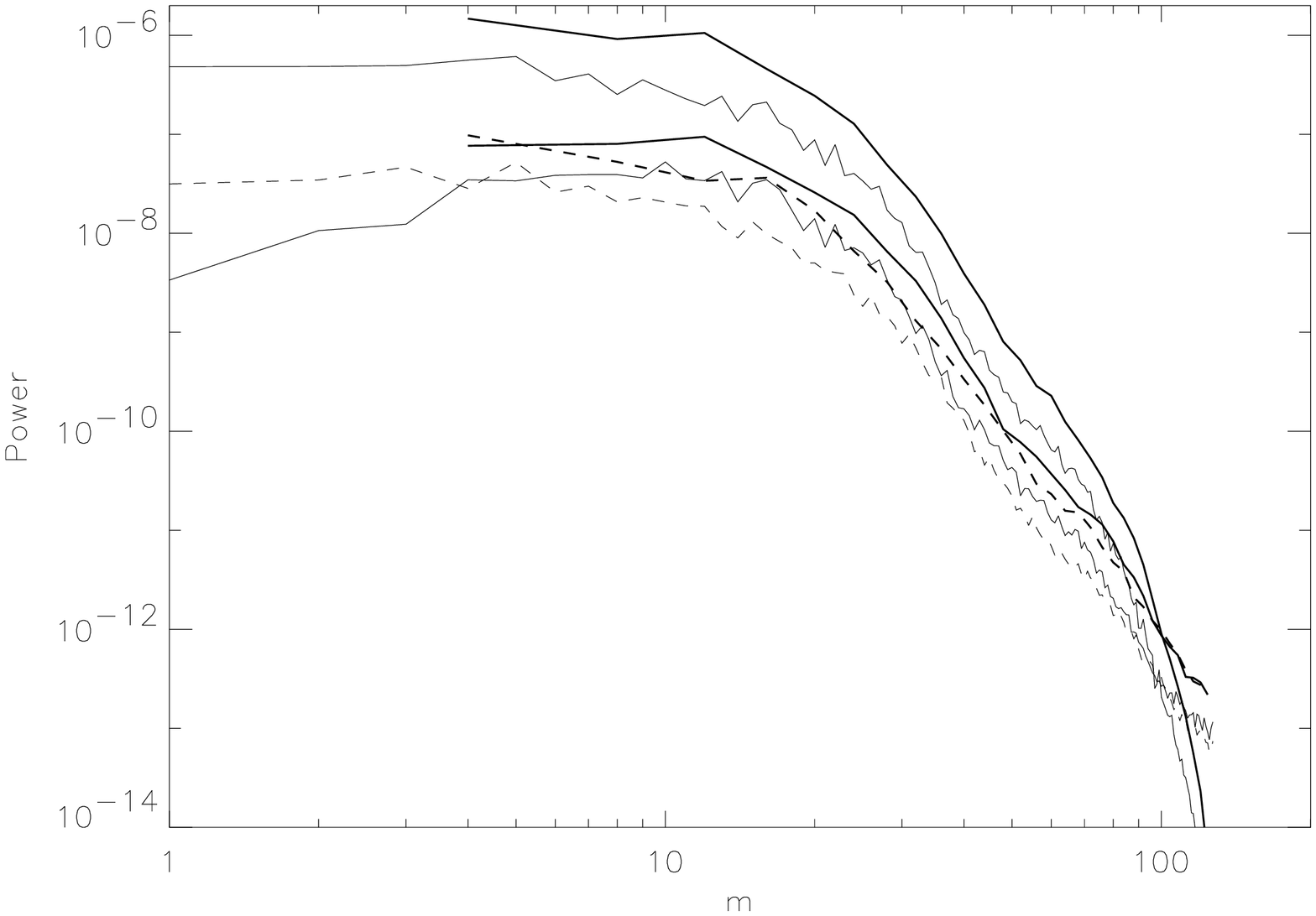,width=3.5in,angle=90}}
\figcaption{Azimuthal fourier power spectra of (a) the
vertically-averaged density, and (b) the magnetic field components
$B_R$ (lower solid curve), $B_\phi$ (top solid curve), and $B_z$
(dashed curve).  The data are taken from at the end time for CK7a (heavy
curves) and the corresponding time in CK7, and averaged over space from
$R=5$ to 20.  Density peaks at the smallest $m$ wavenumbers, 
while the magnetic power spectra have a break in the slope 
between $m=10$ and 20.  In general, there is good qualitative agreement 
between the two simulations over their common angular domain.  
}
\end{figure}

The situation is somewhat different for the magnetic field.  Figure 8b
shows the azimuthal power spectra for the magnetic field components in
runs CK7 and CK7a.  What is most striking here is the break in the
spectral slopes between $m=10$ and 20.  The same phenomenon was noted
by Armitage (1998; see Fig. 3) in a full $2\pi$ simulation of a
cylindrical disk with an initial vertical field.  These plots suggest
CK7 and CK7a are so similar because the primary input from the MRI is
on scales $\sim H$, as would be expected for a vertical field, and $H
\ll R$.  The inverse cascade to smaller $m$ numbers leaves CK7 with
roughly 10\% of the energy at the largest azimuthal scales, but this
doesn't significantly alter the evolution of the MRI.

The azimuthal domain size might be more crucial in simulations of disks
containing an initial toroidal magnetic field rather than a vertical
field.   With toroidal fields the instability depends directly upon the
azimuthal wavenumber $m$.  For example, the most unstable wavenumber
has $m/R v_A \approx \Omega$.   For low $m$ this corresponds to a field
for which $v_A \sim v_\phi$, which would be an exceptionally strong
field.  Under such circumstances a full $2\pi$ global treatment would
obviously be required.  Even with weaker toroidal fields, when $v_A \ll
v_\phi$, low $m$ modes remain unstable, albeit with smaller growth
rates.  Further, in a shearing background, nonaxisymmetric waves have a
time-dependent radial wavenumber.  According to linear theory (Balbus
\& Hawley 1992; Terquem \& Papaloizou 1996) 
for a given azimuthal mode $m$, maximum field
amplification occurs  when the the wavenumber quantity $(k/k_z)^2$ is
small, hence field growth is maximized with low values $k_R$ and/or
high values of $k_z$.  Since $k_R$ evolves as $mt d\Omega / dR$ a
reduction in total amplification can occur if low $m$ modes are removed
by a reduced angular domain size.

For a test comparison, we examine two cylindrical disk simulations that
were computed for a separate project.  These simulations are labeled
NK1 and NK1a and are listed in Table 1.  Each begins with identical
initial conditions except that NK1a uses a $\pi/2$ angular domain and
NK1 the full $2\pi$.  These simulations use a Newtonian gravitational
potential with $GM=1$ on a grid that runs from $R=0.25$ to 3.75, and 0
to 0.2 in $z$.  Note that this make the time units different from the
other runs; the orbital period at the center of the grid ($R=2$) is
17.8.  The initial condition consists of an isothermal Keplerian disk.
The temperature profile is fixed so that the Mach number ${\mathcal M}
= c_s/v_\phi = 20$ is constant with radius.  The pressure is
also constant, and the density increases as $R$.  The initial magnetic
field is toroidal and of strength $\beta=4$.  The most unstable
wavenumber is $m\cong v_\phi/v_A= {\mathcal M}\sqrt{2 \beta} =57$; 
this is well above the $m=4$ limit of the $\pi/2$ grid.

Both simulations have a similar local growth rate for the perturbed
magnetic field during the first few local orbital periods.  Then the growth
rate for NK1a drops off while NK1 continues unabated to  saturation at a
time of 10 local orbits.  The field energy in NK1 is higher than in
NK1a, but NK1a continues to grow
with a slower rate, and manages to achieve about the same level as NK1
after 20 orbits.  Beyond this point in time, the energies fluctuate
around a mean, and the mean values differ by about 10\%.  Similarly,
the mean magnetic stress in NK1a is 13\% lower than in NK1.  

An examination of the fourier power spectrum for the density and the
magnetic field components in run NK1 shows that the power in the field
components continues to rise until about $m=4$.  Beyond, to smaller $m$
values, the power is flat ($B_\phi$) or slightly decreasing ($B_R$).
In contrast, the power in CK7 turns over at larger $m$ values.  The
difference is that when the background field is toroidal there is power
at $m=0$.  Further, the toroidal field is unstable for all nonzero $m$ below
the critical wavenumber $m= {\sqrt 3} v_\phi/v_A$.

To conclude, although there are significant differences in the initial
growth stage of the MRI, and there is power in the lowest $m$ modes, by
the end of the run the reduced domain size has the same effect on the
toroidal field simulations as for those with vertical field:  a 10\%
reduction in averaged field strengths.  The reduction of
the $\phi$ domain is a good approximation because the weak-field MRI is
essentially a {\it local} instability.  The MHD turbulence is driven
locally, and the qualitative behavior of the disk is unchanged so
long as the field remains weak, $\beta > 1$.

\subsection{Equation of State}

With the simple energy equation employed in these simulations there are
two interesting limits:  the adiabatic and the isothermal equations of
state.  Over the long term these two equations of state should 
lead to divergent evolutions as the adiabatic disk heats up.
The present comparison has the modest goal of investigating
whether the equation of state has an impact over a relatively brief
period of initial evolution in the cylindrical disk.

Run CK7b is a repeat of CK7a using an isothermal equation of state.
The initial linear growth phase in the two runs is essentially
identical.  Once turbulence sets in the runs vary in detail, but not in
any systematic way.  In particular there are no significant differences
between the two runs in terms of stress at $r_{ms}$ or the change in
$\ell$ between $r_{ms}$ and the inner boundary.  Although the accretion
rates in both runs differ from each other, they both also vary strongly
in time, and the average difference between the two runs is less than
the fluctuations level seen in either run alone.

The main differences observed in these two runs are consistent with
what would be expected for the these equations of state.  In CK7a
the temperature varies with radius, rising over the course of the
simulation by as much as 30\% inside of $R=10$, while falling rapidly
inside of $r_{ms}$.  The temperature has also declined slightly outside
of $R=25$ due to expansion of the disk off the outer boundary.  The
density fluctuations (\ref{fluctuation}) in the isothermal run CK7b are
larger on average over the whole disk:  the mean $\delta \rho/\rho$ is
0.73 versus 0.60 in CK7a.

\section{A Hydrodynamic Disk}

Cold disks such as protoplanetary disks may lack sufficient ionization
to couple to the magnetic fields.  What happens in a purely
hydrodynamic disk?  Since turbulence, angular momentum transport, and
net accretion in disks result from the action of the MRI, evidence to
date suggests that very little happens in such hydrodynamic disks.
There is now a growing body of simulations (Stone \& Balbus 1996;
Balbus, Hawley \& Stone 1996; Hawley, Balbus \& Winters 1999; Godon \&
Livio 1999a) which are consistent with the conclusion that
differentially rotating disks are hydrodynamically stable to both
linear and finite-amplitude local perturbations.  Although there is no
particular reason to believe that {\it local} hydrodynamic stability
properties would be altered in a fully global disk, the hydrodynamic
disk nevertheless remains an important limiting test case to consider.

To follow the evolution of an initially turbulent disk in the {\it
absence} of Lorentz forces, the output from CK5 at time $t=3450$ is
evolved forward in time (to $t=4850$) purely hydrodynamically (run
HK5).  It is perhaps a bit self-inconsistent to ask how turbulence that
is {\it generated} by magnetic fields subsequently evolves 
hydrodynamically.  Of course when there is no self-consistent 
hydrodynamic turbulence, and one wishes to observe the hydrodynamic 
decay of turbulence, it is necessary to initialize the turbulence 
one way or another.

In any event, the results from HK5 are not surprising.  After a brief
period of readjustment to the loss of magnetic tension and pressure
forces, the turbulent kinetic energies drop (Fig. 9).  In particular,
the vertical kinetic energy $1/2 \rho v_z^2$ declines exponentially and
the system became increasingly $z$-independent.  This is consistent
with the cylindrical limit and the Taylor-Proudman theorem which holds
that in a steady, inviscid flow, slow motions in a rotating fluid
should be two-dimensional.  All accretion stops, and the inner edge of
the disk readjusts slightly, moving outside of $R=4$.  Consistent with
past local hydrodynamic simulations, there is no evidence for sustained
turbulence or the development of any local hydrodynamic instability.
Nor is there evidence for a strongly  growing global mode.

Inward and outward propagating pressure waves are present throughout
the simulation.  These are in the form of tightly-wrapped trailing
spiral waves, and they persist with constant or diminishing amplitude.
A comparison of the azimuthal fourier power spectrum for density from
CK5 and HK5 finds that CK5 has more power for all wavenumbers $m$.  The
ratio of power in CK5 to HK5 rises from 2:1 at $m=4$ (lowest
wavenumber) up to 34:1 for $m=44$.  The high-$m$ power in CK5 is driven
by the MRI; its amplitude drops promptly with the elimination of the
magnetic terms in HK5.  The dominant $m=4$ waves of HK5 appear to be
driven mainly from the inner edge of the disk at frequencies close to
the local orbital frequency there.  The inner edge of the disk has a
nonaxisymmetric structure that resembles a surface wave.  This type of
wave is seen to arise in Keplerian disks with a reflective inner
boundary, and it is an example of a Papaloizou-Pringle instability
(Godon \& Livio 1999b).  The pressure waves can, in principle, drive a
small net inward drift within the disk.  If one averages the accretion
rate at each radius over the entire evolution there is a slight average
inward accretion in the inner portion of the disk amounting to no more
than $\dot M \sim 0.001$.  A similarly averaged accretion rate in the
magnetic disk is 0.075.

\begin{figure}
\centerline{\psfig{file=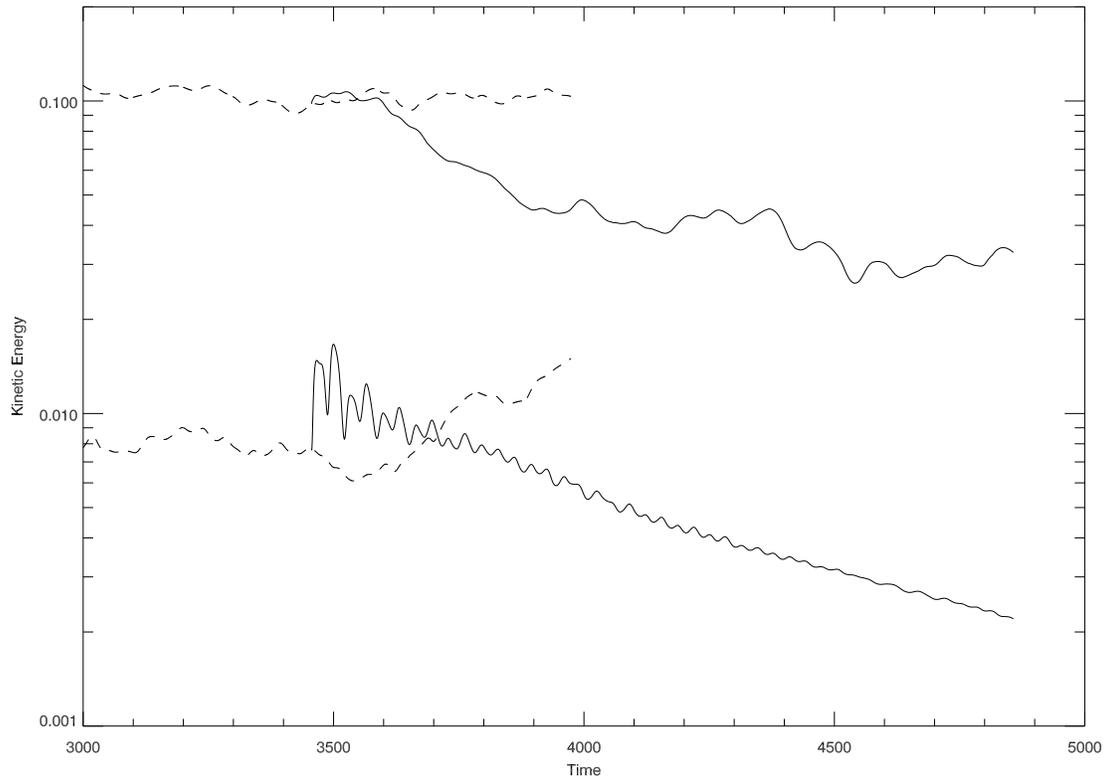,width=6.0in,angle=90}}
\figcaption{Time-history of the radial and vertical kinetic energy 
in MHD run CK5 (dashed line) and hydrodynamical run HK5 (solid
line).  HK5 begins from the data of CK5 at time $t= 3450$.  After an
initial readjustment to the loss of magnetic force, the kinetic
energies in the hydrodynamic run decay away.
}
\end{figure}

\section{Discussion}

\subsection{Evolution of Keplerian disks: local versus global}

The cylindrical disk is global in radius and azimuthal angle, but
essentially local in $z$; it represents the next step beyond local
shearing box simulations.  The first cylindrical disk was computed by
Armitage (1998), and two cylindrical Keplerian disk models were
computed in Hawley (2000), but the limited radial extent of the grid in
these models restricted the amount of evolution possible before the
outer boundary significantly affected the simulation.  
ARC computed three cylindrical disks with an initially gaussian
density distribution, contained entirely upon the grid.
They examine the inflow through the marginally stable orbit in a
pseudo-Newtonian potential with an emphasis on the stress 
there.  In this paper we consider the evolution of an MHD turbulent
Keplerian cylindrical disks from simple initial conditions, but with a
much larger radial extent.

First, how well do local shearing box simulations describe the state of
the instability, turbulence, and transport due to the MRI when compared
to global models?  The answer seems to be that local models do quite
well, as long as the questions being asked of them are appropriate to
the local approximation.  Even in a global disk the MRI is a local
instability; the wavelengths of the fastest growing modes are always
less than $H$ and $R$ so long as the magnetic field is weak.  The great
similarity between the $\pi/2$ and $2\pi$ CK7 models illustrates this
principle.

One of the major results from the local simulations is the importance
of the background field topology.  Local simulations (HGB95) find that
a net vertical field leads to greater amplification of the initial
field compared to a simulation beginning with a purely toroidal field.
Stronger field amplification can lead to stronger turbulence and
greater stresses.  A comparison between models CK5 and CK6, and between
as the vertical and toroidal simulations of ARC, support these
conclusions in the global context as well.  The ``efficiency'' of the
Maxwell stress is measured by $\alpha_{mag}$, the ratio of the Maxwell
stress to the magnetic pressure.  When the toroidal and radial field
energies are comparable and the fields fully correlated to produce
stress with the correct sign to transport angular momentum outward,
$\alpha_{mag} \sim 1$.  Simulations show that in both the global
and local systems the turbulent state is dominated by toroidal field;
this is particularly true when the initial background field is
toroidal.  With vertical initial fields the toroidal and radial fields
are more comparable (although the toroidal field energy remains the
largest).  This is reflected in the values of $\alpha_{mag}$:  for
vertical fields $\alpha_{mag} \approx 0.5$, while toroidal fields
have a value closer to 0.3.

It should be mentioned that in the local simulations it is
straightforward to measure the Reynolds stress, $\rho \delta v_r \delta
v_\phi$; there is a well defined
background shearing rate which allows an unambiguous definition of
$\delta v_\phi$ as well as a limited volume over which to
average.  It is much more difficult to do this in the global
simulations since at any given moment the background flow exhibits
substantial deviations from, say, a Keplerian value.  In local
simulations, however, the Maxwell stress always dominates over the
Reynolds stress by a factor of several.  While measuring only the
Maxwell stress for global simulations provides only a lower limit,
$M^{R\phi}$ should nevertheless account for the majority of the
stress.

Although the MRI is local and  many properties of the resulting MHD
turbulence are local as well, the stress is proportional to $P_{mag}$,
the saturation amplitude of the field, and this
might well be determined by global properties such as the scale height $H$
or the ratio $H/R$.  So far this has been difficult to assess.
The traditional Shakura-Sunyaev $\alpha$ parameter
is set by the relation $\alpha = \alpha_{mag}/\beta$.  Thus $\alpha =
0.1$--$0.01$ requires $\beta \sim 3$--50.  In local simulations
vertical fields tend to saturate near the lower end of this $\beta$
range, and toroidal fields at the upper end, unless the toroidal field
{\it began} with $\beta \sim $1--10.  If one considers the current
global simulation results, both from this paper and from previous cited
works, the impression is that global simulations produce lower $\beta$
at saturation and larger $\alpha$ values than the local models.
However, one should be cautious drawing a general conclusion at this
stage.  The global simulations have initial vertical fields (or
poloidal field loops), or initially strong toroidal fields.  For such
fields the resulting saturation levels near $\alpha \sim 0.1$ are fully
consistent with the local simulations.  Local simulations saturate at
higher $\beta$ values when the initial field consists of weak random
field, or weak toroidal field, and these cases have not yet been
investigated globally, in part due to the higher resolution required.

One aspect that emerges from these and other global simulations is the
difficulty of characterizing a ``steady state'' disk.  In
the global simulations all quantities vary strongly both in time and in
space. Although one can average over space and over many orbits and
obtain relatively smooth spatial distributions, significant
fluctuations are always present.

Absent from the local models, but present in global disks are effects
such as a net accretion and spiral wave propagation.  The MRI is
inherently time unsteady and produces fluctuations in the disk at
frequencies close to the local orbital frequency, generating
magnetoacoustic waves.  It not surprising that strong MHD turbulence
should generate such magnetoacoustic waves.  Blaes and Balbus (1994)
show that the presence of toroidal fields couples the compressible and
incompressible modes of the MRI, and when $\beta$ approaches order
unity, significant acoustic modes are expected.  In These spiral waves
are generated at small radii and propagate out through the full radial
extent of the disk.  This is one way that the turbulence at the inner
part of the disk exerts a global effect on the disk.

\subsection{Viscous Instability}

A prominent feature of model CK6 at late time is the presence of a
dense ring in the inner region and low density gaps that have formed
within the disk.  These gaps arise due to a type of 
``viscous'' instability, specifically an inverse relationship between the
Maxwell stress and the density.  Although these rings and gaps are most
prominent in CK6, there is evidence for a similar tendency in CK5, the
toroidal field simulation. 

The dense rings that form in CK6 are likely to be enhanced by the
cylindrical approximation.   The rings {\it per se} might not form in
simulations with vertical stratification since the strong magnetic
field located between the dense rings would be buoyant.  Cylindrical
symmetry eliminates the possibility of buoyancy as well as any
development of magnetized outflows.  The tendency for greater stress in
regions of lower density is not due to the cylindrical limit, however,
but due to the increased Alfv\'en speeds in such regions.  The probable
outcome of this tendency in stratified disks will be larger stresses in
the lower density region above the equatorial plane (see, for example,
in the simulations of Miller \& Stone 2000), and the creation of local
regions of strong toroidal field that will be ejected into a corona.
In fact, strong toroidal fields rising out of the disk are a common
feature of the global thick torus simulations of Hawley (2000), HK, and
Machida et al.~(2000).  Another consequence is that the accretion rate
through the disk will be inherently unsteady, and spatially
inhomogeneous.  The two-dimensional axisymmetric global MHD simulations
of Stone \& Pringle (2000), for example, provide evidence of just such
strong radial inhomogeneities in accretion rate and density.

How generic is this tendency to develop a viscous instability?
Interestingly there is no evidence for the formation of dense rings in
CK7, in contrast to CK6.  The most obvious systematic differences
between CK6 and CK7 are the domain size ($\pi/2$ versus $2\pi$), the
equation of state (isothermal versus adiabatic), the absence or
presence of net accretion into the central hole, and total evolution
time.  The influence of the $\phi$ domain seems too small to account
for the difference.  Although an isothermal equation of state can
create greater density fluctuations,  a direct comparison between
adiabatic and isothermal models CK7a and CK7b does not find any
systematic effect due to the choice of equation of state.  Further, ARC
used an isothermal equation of state and did not observe any tendency
toward the formation of dense rings.

This leaves the influence of the inflow through the marginally stable
orbit and the duration of the simulation.  In models that do not
accrete through the marginally stable orbit, mass tends to pile up at
the inner edge of the disk.  The accretion through $r_{ms}$ prevents
this, and creates an inward pointing pressure gradient out through the
disk to radii that are several times $r_{ms}$.  This appears to inhibit
the development of the viscous instability, at least in the inner disk
over the time scales simulated.  The viscous instability might yet
manifest itself in the disk at larger radii, but CK7 was not run far
enough in time for this to occur.  Further simulations and analysis
would be required to test this conjecture further.

\subsection{Stress at the Marginally Stable Orbit}

Since the pseudo-Newtonian potential is used for these simulations, we
can examine the evolution of the stress and specific angular momentum
in the region of the marginally stable orbit.  The cylindrical
simulations demonstrate that the stress is continuous at $r_{ms}$;
indeed, there is nothing special at the precise location of $r_{ms}$ in
any disk quantity.  In the present simulations the slope of the
specific angular momentum $d\ell / dR$ inside the marginally stable
orbit is smaller than seen in the fully global thick disk simulations
of Hawley (2000) or HK.  The slope is close to that reported by ARC
from their simulations of cylindrical Keplerian disks.  What then
determines the degree to which $\ell$ is reduced inside of $r_{ms}$?
In each case there is some nonzero
stress inside of $r_{ms}$.  Naturally, larger stresses have a larger
effect.  The question becomes what circumstances produce those larger
stresses?

A reduced angular computational domain lowers the observed
stress levels, but only by about 10\%.  The use of an
isothermal versus adiabatic equation of state apparently has even less
influence, at least when the temperatures at $r_{ms}$ are comparable.
The internal pressure may have a greater influence, by increasing the
radial distance inside of $r_{ms}$ where the flow remains subsonic,
however there is only circumstantial evidence for this in the results
to date.  The simulations run here were about half as hot as those of
HK, and comparable to the fiducial runs of ARC.  ARC also ran a
simulation with the sound speed cut in half but did not report any
significant differences in the results.

Possibly the most important approximation is the use of cylindrical
geometry in lieu of full stratification.  Stratified simulations find
that the largest magnetic field strengths and Alfv\'en speeds occur in
the lower density regions surrounding the equator (Miller \& Stone
2000).  In the thick disk simulation of HK the specific angular
momentum was nearly constant inside of $r_{ms}$ along the equator.  The
strongest fields and stresses were located above and below the equator,
and this is where the greatest reduction in $\ell$ occurred.  To
investigate this further there appears to be no substitute for
stratified global simulations.

\subsection{Hydrodynamic Stability}

Repeated local simulations have failed to find any evidence for a
purely hydrodynamic local nonlinear instability.  The absence of
magnetic fields has always resulted in the decay of any imposed
turbulence, leading to the conclusion that there is no turbulent
$\alpha$ in unmagnetized disks.  While there are known global
instabilities in hydrodynamic disks such as the Papaloizou-Pringle
instability, or local violations of the H\o iland criteria,  the
circumstances under which these manifest themselves are known, and
these instabilities are unlikely to be generally significant.

As a control a hydrodynamic cylindrical disk is computed.  This disk
begins with a Keplerian disk that is MHD turbulent.  The simulation
consists of turning off the magnetic forces and following the
subsequent hydrodynamic evolution.  The outcome is straightforward:
the disk turbulence dies out promptly.  All vertical structure decays
away and the disk becomes two dimensional.  Global spiral waves
continue to propagate, but their amplitudes are reduced from those seen
with active MHD, particularly for larger azimuthal wavenumbers.  

\subsection{Coherent structures}

Another issue of interest in disks, hydrodynamic or
magnetohydrodynamic, is the possible formation of coherent structures.
For example, Abramowicz et al. (1992) suggested that coherent vortices
might produce observable modulations in the luminosity of disks in
active galaxies.  More recently, vortices have been proposed as sites
for planetesimal formation in protoplanetary disks.

Two-dimensional $(R,\phi)$ hydrodynamic simulations of Keplerian disks
have been carried out by Nauta \& T\'oth (1998), Nauta (1999), and
Godon \& Livio (1999b; 2000).  The results to date can be summarized
thusly:  (1) vortices placed in hydrodynamic Keplerian disks can
survive for many orbits, and (2) vortices have not been observed to
arise spontaneously in initially Keplerian disks.  Point (1) is quite
understandable since the counter-rotating epicyclic flow is an
equilibrium solution.  Point (2) follows from the stability of a
hydrodynamic Keplerian flow; there is nothing to drive the transition.

These points are illustrated by an example where
coherent hydrodynamic structures {\it were} observed to
arise in simulations of globally unstable disks.  This 
is the counter-rotating coherent ``planet'' discovered by Hawley
(1987) in simulations of the nonlinear saturation of the Papaloizou \&
Pringle (1984) instability in slender tori.  Planets form in thick
accretion tori as well (Hawley 1991), although in a less dramatic
fashion than in the slender torus limit.  These structures were
described analytically  by Goodman, Narayan, \& Goldreich  (1987) as
fluid undergoing elliptical counter-rotating epicyclic motion.  In
these planets the Coriolis force is in balance with the fluid pressure,
creating an equilibrium structure.  These structures can be described
as vortices, but the essence of their nature is epicyclic motion.
While they are a local equilibrium solution for a differentially
rotating fluid, it is unclear when such solutions can arise from an
initially axisymmetric disk flow.  In the slender torus, the epicyclic
planets develop through the action of the Papaloizou-Pringle
instability.  The instability triggers the transition from one
equilibrium to the other.  The prospects for such a transition in a
stable Keplerian disk seem less promising.

In run HK5 there is no sign of emerging or sustained coherent
structures such as counter-rotating epicyclic flows.  When looking for
coherent structures, however, it is hardly necessary to confine one's
attention to hydrodynamic disks.  We can also ask whether
self-sustained MHD turbulence generates coherent structures.  No such
structures have been observed in local simulations (e.g., Brandenburg
et al.  1995).  Similarly, in global MHD simulations, cylindrical or
otherwise, there has been no evidence for the formation of
coherent structures.

It is worth reiterating that the coherent vortices, i.e., the
hydrodynamic planet structures, owe their existence to the presence of
stable epicycles, but it is precisely those epicycles that are
disrupted by the action of the MRI.  Still, analytic MHD equilibrium
coherent structures have been found by Balbus \& Ricotti (1999) in the
local limit  which are magnetic analogues to the hydrodynamic planet
solution.  It has yet to be demonstrated that these solutions are
stable, or that they can develop in an MHD disk.

\subsection{Impact of numerical model assumptions}

Given the demands of three dimensional simulations, it is useful to
explore ways in which the global problem can be restricted without too
much loss of significance.  One such simplification is the cylindrical
disk limit, and another is the restriction to a smaller domain in
$\phi$.  This paper has examined the influence of the later
approximation for a model with an initial vertical field, and one with
an initial toroidal field.

The explicit comparison between simulations using the full $2\pi$ and
ones that span only $\pi/2$ in angle suggests that the restricted
azimuthal domain is indeed a useful approximation.  The vertical field
simulations are qualitatively very similar throughout.  Although the
initial growth stage in the toroidal field cases are noticeably
different in the two different computational domains, the final
turbulent states are again very similar.  With vertical fields the
magnetic azimuthal power spectrum has a significant break at relatively
high $m$ wavenumbers, consistent with the fastest growing modes of the
vertical MRI.  With toroidal fields the power spectra level out toward
low $m$ and do not show the same break.  Again, this is consistent with
the MRI:  weak toroidal fields are unstable for all $m$ less than
$\approx v_\phi/v_A$.  In either case, however, the power spectra of
the $2\pi$ and $\pi/2$ simulations are very similar where their
wavenumbers overlap.  Total stress and accretion rate values are
reduced by about 10\% in the restricted domains, indicating that 10\%
of the energy is to be found on the largest azimuthal scales.  This is
not entirely negligible, but the qualitative difference
does not appear to be profound.

A significant factor in global simulations remains the grid resolution
which will always fall short of what one can achieve with a local
simulation.  Here the average radial grid zone size is $\Delta R =
0.23$.  The standard radial resolution in the local simulations of
HGB95 was almost a factor of 10 better.  Because of the lack of
stratification, the vertical resolution in the cylindrical simulations
is comparable to the local model.  A resolution comparison done by HK
found larger magnetic energies with better resolution, a result
consistent with more detailed resolution studies in the local shearing
box.  Even with the grid resolution that is presently possible,
however, global simulations generate and maintain MHD turbulence with
physically significant stress levels.

\section{Conclusion}

Cylindrical disk simulations are a useful tool for investigating global
evolution of disks evolving due to magnetically driven turbulence.
Such simulations demonstrate that the conclusions developed in
the local shearing box model hold in the global context as well.  As in
the local model the MRI grows rapidly and produces MHD turbulence with
a significant Maxwell stress.  The turbulence is more vigorous and more
efficient in producing stress for a given total magnetic pressure when
driven by an initial field that is vertical rather than toroidal.
Hydrodynamics alone seems no more effective at creating or sustaining
turbulence in a global model than it is in a local one.

In addition to reaffirming local properties of the MRI, cylindrical
disk simulations illuminate disk characteristics that are truly
global.  A net accretion rate is one such property, but there are also
important nonlocal structural features.  Tightly-wrapped low-$m$ spiral
waves are prominent.  The final accretion through the marginally stable
orbit provides an example of a highly nonaxisymmetric spiral
flow.  Particularly interesting in the present simulations is the
tendency for radial variations in Maxwell stress to concentrate gas
into rings, creating substantial spatial inhomogeneities.

A perennial question is the degree to which these simulations resemble
traditional steady state $\alpha$-disk models.  They do in so far as
they accrete in direct response to internal stress, specifically due to
MHD turbulence.  Beyond that, however, there are significant
differences.  The simulations are characterized by large scale
variability in space and time in all variables.  The stress is
proportional to the magnetic pressure which is itself only indirectly
related to other disk parameters.   In the simulations it is possible
to approximate a quasi-steady state only with broad-stroke averages.
In part this is due to the initial conditions (e.g., isolated tori or
constant density slabs) which are far from a possible accreting steady
state solution.  To address this issue it will be useful to attempt
simulations that begin with more realistic initial states.  Results to
date indicate, however, that analytic disk models are likely to prove
woefully inadequate in describing detailed spatial and temporal disk
properties so long as they are based upon a strict $\alpha$ formulation
with $\alpha$ a constant in space and/or in time.

Although cylindrical disk simulations provide a valuable point of
reference for future work, the lack of vertical stratification is
clearly a major limitation for investigating many important
physical processes.  This is obviously true for the development of a
magnetized corona, or the launching of winds or jets.  It appears also
to be an important factor in measuring the stress in the disk at the
marginally stable orbit.  Stratified global thin disk simulations are
the next logical step to contrast with existing global thick disk
models.  Stratified thin simulations will require far more vertical
grid zones centered around the equator than are used in cylindrical
disks.  However, the tests presented here suggest that a reduction in
the $\phi$ domain is a acceptable problem simplification, as long as
the potential for some small quantitative reduction in energy levels 
is kept in mind.

\begin{acknowledgments}

I thank Steve Balbus, Julian Krolik, Jim Stone, and Wayne Winters for
useful discussions related to this work.  Wayne Winters supplied data
from his unpublished simulations NK1 and NK1a for the analysis in
\S3.4.  This work was supported by NSF grant AST-0070979, and NASA
grants NAG5-9266 and NAG5-7500.  Simulations were carried out on the
Cray T3E and T90 systems of the San Diego Supercomputer Center of the
National Partnership for Advanced Computational Infrastructure, funded
by the NSF.

\end{acknowledgments}

\clearpage

\begin{center}
{\bf TABLE} \\ CYLINDRICAL KEPLERIAN DISK SIMULATIONS \\

\addvspace {0.5cm}

\begin{tabular}{cccccc} \hline \hline
\\ [-0.3cm]
Model&
$(R,\phi,z)$ Domain&
Grid &
Equation of State&
Initial Field  &
End time \\
\hline
\\ [-0.2cm]

CK5& 1.5-61.5, $\pi/2$, 2.0& $256\times 58\times 24$ & isothermal 
& toroidal & 3972 \\
HK5& 1.5-61.5, $\pi/2$, 2.0& $256\times 58\times 24$ & isothermal 
& hydro & 1400  \\
CK6& 1.5-61.5, $\pi/2$, 2.0& $256\times 64\times 32$ & isothermal 
& vertical& 4097  \\
CK7& 1.5-61.5, $2\pi$, 0.8& $256\times 256\times 32$ & adiabatic 
& vertical & 2575  \\
CK7a& 1.5-61.5, $\pi/2$, 0.8& $256\times 64\times 32$ & adiabatic 
& vertical & 1500  \\
CK7b& 1.5-61.5, $\pi/2$, 0.8& $256\times 64\times 32$ & isothermal
& vertical & 1500  \\
NK1& 0.25-3.75, $2\pi$, 0.2& $128\times 128\times 32$ & isothermal 
& toroidal & 71.1 \\
NK1a& 0.25-3.75, $\pi/2$, 0.2& $128\times 32\times 32$ & isothermal 
& toroidal & 66.6 \\

\\
\hline
\end{tabular}
\end{center}

\end{document}